\documentclass[normal]{aa}
\usepackage{psfig}
\def\M21{\hbox{Mrk~421} }
\def\xmm{\hbox{$XMM-Newton$} }
\def\etal{et al.\/ }

\def\ltsima{$\; \buildrel < \over \sim \;$}
\def\simlt{\lower.5ex\hbox{\ltsima}}            % < over ~
\def\gtsima{$\; \buildrel > \over \sim \;$}
\def\simgt{\lower.5ex\hbox{\gtsima}}            % > over ~
\begin{document}
\title{XMM$-$Newton timing mode observations of Mrk 421}
\author{ W. Brinkmann\inst{1}, I. E. Papadakis\inst{2,3}, 
     C.Raeth\inst{1}, P. Mimica\inst{4},  F. Haberl\inst{1} }
%\offprints{I.E. Papadakis; e-mail: jhep@physics.uoc.gr}
\offprints{W. Brinkmann; e-mail: wpb@mpe.mpg.de}
\institute{Max--Planck--Institut f\"ur extraterrestrische Physik,
  Postfach 1312, D-85741 Garching, FRG
\and IESL, FORTH, 711 10 Heraklion, Crete, Greece
\and Physics Department, University of Crete,
    710 03 Heraklion, Crete, Greece  
\and  Max--Planck--Institut f\"ur Astrophysik,
    Postfach 1317, D-85741 Garching, FRG}
\date{Received ?; accepted ?}
 \abstract{
 We present  the results of a detailed temporal analysis of
 the bright BL Lac object Mrk~421 using  the three available long
 timing mode observations by  the EPIC PN camera.  This detector mode
is characterized by its long life time and is largely free
of photon pile-up problems. The source was found in
different intensity and variability states differing by up to more than
a factor of three in count rate. A time resolved cross 
correlation analysis between the soft and hard energy bands revealed 
that the characteristics of the correlated emission, with lags of both
signs, change on time scales of a few $10^3$ seconds. 
Individual spectra, resolved on time scales of $\sim$ 100\,s,
can be quite well fitted by a broken power law and we
find significant spectral variations on time scales
as short as $\sim 500-1000$ sec.
Both the hard and the soft band spectral indices show a non-linear
correlation with the source flux.  A simple power law model
of the form $\Gamma\propto {\mathrm {flux}}^{-a}$ with $a_{\mathrm {hard}}\sim 0.13$ and
$a_{\mathrm {soft}}\sim 0.22$ describes rather well the observed trend of decreasing
 $\Gamma$ values with increasing flux, which appear to ``saturate" at the same limiting
value of $\Gamma\sim 1.8$ at the highest flux levels. 
A comparison of the observed light curves with numerical results
from relativistic hydrodynamic computer simulations of the currently
favored shock-in-jet models indicates that any determination of the jet's 
physical parameters from `simple' emission models must be 
regarded with caution:  at any time we are seeing the emission from
several emission regions distinct in space and time, which are 
connected by the complex hydrodynamic evolution of the non-uniform jet.
  }
\titlerunning{XMM$-$Newton timing analysis of Mrk 421}
\authorrunning{W. Brinkmann et al.}
\maketitle

\section{Introduction}
\smallskip
BL Lacs are radio-loud AGN dominated by relativistic jets seen at small
angles to the line of sight (\cite{UP95}). They are characterized by large
luminosities, large and variable polarization, and strong variability. A
substantial part of the information we possess about these objects is
obtained from the analysis of temporal variations of the emission and the
combined spectral and temporal information can greatly constrain the jet
physics. Time scales are related to the crossing time of the emission
regions which depend on wavelength and/or the time scales of the relevant
physical processes. The measured lags between the light curves at
different energies as well as spectral changes during intensity variations
allow us to probe the physics of particle acceleration and radiation in the
jet.
 
Mrk~421 is the brightest BL Lac object at X-ray and UV wavelengths and it
is the first extragalactic source discovered at TeV energies (Punch \etal
1992). The nearby (z = 0.031) object was observed by almost all
previous X-ray missions and shows remarkable X-ray variability correlated
with strong activity at TeV energies (e.g., Takahashi \etal 1996, Maraschi
\etal 1999). Structure function and power density spectrum analyses
indicate a roll-over with a time scale of the order of 1 day or longer
(Cagnoni \etal 2001, Kataoka \etal 2001) which seems to be the time scale
of the successive flare events. On shorter time scales less power 
in the variability is found with a steep slope of the power density
spectrum $\sim f^{- (2 - 3)}$(Tanihata 2002, Brinkmann et al. 2003).
  
The smooth featureless continuum spectra of \M21\ from the radio to the
X-ray band imply that the emission in these energy bands is due to
relativistic particles radiating via the Synchrotron process; the hard
X-rays and $\gamma$-rays are likely to be produced by inverse Compton
scattering of synchrotron photons by the same electrons (Ulrich et al.
1997 and references therein). Broken power law models provided better fits
to the ASCA 0.6$-$7 keV data than a simple power law (Takahashi \etal
1996) with a break energy of $\sim$ 1.5 keV, and the change of the power
law index at the break point is $\Delta \Gamma \sim 0.5$, i.e., the
spectrum is steeper at higher energies.

With the wider energy range of BeppoSAX it became clear that the simple
models are not adequate descriptions of the downward curved Synchrotron
spectra (Fossati \etal 2000b) and continuously curved shapes had to be
employed (Inoue \& Takahara 1996, Tavecchio \etal 1998). The Synchrotron
peak energy varied between 0.4$-$ 1~keV, the spectral index at an energy
of 5~keV between $2.5 \leq \Gamma \leq 3.2$. Both quantities are
correlated with the X-ray flux: with increasing flux the synchrotron peak
shifts to higher energies and the spectrum gets flatter at higher
energies.

The emission of the soft X-rays is generally well correlated with that of
the hard X-rays and lags it by 3$-$4~ksec (Takahashi \etal 1996, 2000,
Zhang \etal 1999, Malizia \etal 2000, Kataoka \etal 2000, Fossati \etal
2000a, Cui 2004), however, significant lags of both signs were detected
from several flares (Tanihata \etal 2001, Takahashi et al. 2000).
  
These results were obtained from data with relatively low signal-to-noise
ratio, integrated over wide time intervals (typically one satellite orbit)  
and by analyzing data from prominent flares with time scales of a day.
Recently, \xmm with its high sensitivity, spectral resolving power and
broad energy band provided uninterrupted data with high temporal and
spectral resolution. In an analysis of the early \xmm data (Brinkmann
\etal 2001) for the first time the evolution of intensity variations could
be resolved on time scales of $\sim$ 100 s. Temporal variations by a
factor of three at highest X-ray energies were accompanied by complex
spectral variations with only a small time lag of $\tau =
265^{+116}_{-102}$ s between the hard and soft photons. Sembay \etal
(2002) confirmed these short lags in an analysis of a larger set of \xmm
observations of Mrk 421 and they show that the source exhibits a rather
complex and irregular variability pattern - both temporarily and
spectrally.  From \xmm observations of PKS 2155-304 Edelson \etal (2001)
suggest that previous claims of soft lags with time scales of $\sim $
hours might be an artifact of the periodic interruptions of the low-Earth
orbits of the satellites every $\sim$ 1.6 hours. This claim was questioned
by Zhang et al. (2004) who show that, although periodic gaps introduce
larger uncertainties than present in evenly sampled data, lags on time
scales of hours cannot be the result of periodic gaps.
 
In an extended analysis of all available \xmm PN observations of \M21\
Brinkmann et al. (2003) split up the individual light curves into shorter
sub-intervals in which the flux appeared to show some similar specific
behavior. The cross-correlation analysis of these individual soft and hard
band light curves demonstrated that they are in general well correlated at
zero lags, but sometimes the hard band variations lead the soft band
variations by typically $\sim$ 5 min, in two cases the soft band lead the
hard band variations.  The delays appear to be correlated with a harder
spectrum during higher intensities. Recently, Ravasio et al. (2004) used
three \xmm observations in Nov/Dec. 2002 when \M21\ was highly variable.  
During two large flares the source showed strong spectral evolution and
temporal lags between the different energy band variations of both signs,
with lags of up to more than 1000 secs.
 
\setcounter{table}{0}
\begin{table*}[t]
\small
\tabcolsep1ex
\caption{\label{data} Details of the observations }
\begin{tabular}{llcccccc}
\noalign{\smallskip} \hline \noalign{\smallskip}
\multicolumn{1}{c}{Orbit} & \multicolumn{1}{c}{Observing date} &
\multicolumn{1}{c}{PN mode} &
\multicolumn{1}{c}{Filter} & \multicolumn{1}{c}{Duration} &
\multicolumn{1}{c}{}Live Time & \multicolumn{1}{c}{count rate} &
\multicolumn{1}{c}{source count rate} \\
\multicolumn{1}{c}{  } & \multicolumn{1}{c}{(UT)} &
\multicolumn{1}{c}{ } &
\multicolumn{1}{c}{ } & \multicolumn{1}{c}{[ ks ]} &
\multicolumn{1}{c}{[ ks ]} & \multicolumn{1}{c}{on chip} &
\multicolumn{1}{c}{[0.6$-$10 keV]$^{\clubsuit}$} \\
\noalign{\smallskip} \hline \noalign{\smallskip}
 084 & May 25, 2000: 03:53 - 10:11 & Timing & Thick &
   22.4 & 16.78 & 373.3 & 221.6$\pm$0.1\\
 546 & Dec 01, 2002: 23:18 - 18:44$^{\dagger}$ & Timing & Medium &
   71.2  & 59.00 & 303.9 & 176.9$\pm$1.3\\
 807 & May 06, 2004: 02:38 -- 21:11 & Timing & Medium &
 66.7 & 25.63 & 841.3  & 491.7$\pm$0.2\\
\noalign{\smallskip}\hline
\end{tabular}
\medskip

$^{\dagger}$: Next day \\
$^{\clubsuit}$: Background subtracted source count rate for the spectral fit \\
\end{table*}

In this paper we will present a detailed analysis of the three currently
available XMM data sets on \M21\ in which the PN camera was operated in
timing mode. We re-processed the data and used the most recent response
functions. In previous observations of \M21\ even the high time resolution
of the PN in small window mode sometimes proved to be insufficient to
prevent photon pile up when the source was in a high intensity state. This
led to a reduction of the signal to noise ratio for the spectral data and
to frequent unpredictable large observational gaps in the light curves,
due to the extremely high work load of the on-board electronics (FIFO
overflows) and the limits of the telemetry rate (counting mode). Timing
mode data have a very high photon collecting efficiency of 99.5\% of the
frame time and do not suffer from pile-up problems which would, with other
detector modes, lead to a substantial loss of data for the extremely high
count rates of \M21.
 
Two of the data sets, that of orbit 84 as well as the short flare in orbit
546 have been presented previously (Brinkmann et al. 2003 and Ravasio et
al. 2004, respectively), however, only in a more limited analysis and
without special emphasis on their short time scale behavior. The third
dataset, from orbit 807, when the source is extremely bright, has not been
published before.  In our analysis we take full advantage of the excellent
statistical quality of the timing mode data sets and, in particular, we
will for the first time analyse the data on time intervals considerably
shorter than the individual observations as the source exhibits changes in
its spectral properties on time scales down to a few thousand seconds. We
will study the spectral slope variability on time scales as short as $\la
$ 1000 sec and how that correlates with the flux variations; we search the
cross correlation function over very short periods and discuss how it
changes with time as the flux varies.

We will first present the data and demonstrate the spectral complexity of
\M21\ in broad band spectral fits. We will discuss the `standard'
cross-correlation analysis of the data which is commonly used to
investigate the cross-links between the hard and soft band light curves
and will then introduce a sliding-window technique to study the
cross-correlations on shorter time scales. In Sect. 4 we will present a
detailed analysis of the source properties using temporally resolved
spectral analyses.  We then discuss these results in the framework of the
shock-in-jet model with some emphasis on numerical simulations of the
hydrodynamic evolution of non-uniform relativistic jets.  Final
conclusions will be given in Sect. 6.

\section{The XMM-Newton observations}
\smallskip
Mrk~421 was observed repeatedly during the XMM-Newton mission, mostly as a
calibration source for the RGS.  During three of these observations the PN
camera was operated in timing mode, once with a thick filter (orbit 84),
the other times with medium filter.  This mode has the advantage that even
at highest source count rates, pile up, and thus spectral distortions, can
be neglected. The details of the observations are given in Table
~\ref{data}, where we first list the orbit in which the observation took
place, the date of the observation, the instrument settings and the
duration of the observation. Due to the high count rate from \M21 the
detector falls frequently into counting mode, which introduces gaps of
typically 21 s duration into the data stream. This happens when
the read-out buffer is 75\% full, and the rate at which this occurs
depends critically on the work load of the detector. For example, in orbit
807, with the extremely high count rate on the chip, about every 20 sec of
data are followed by this $\sim$ 21 sec gap; the observation consists of
$\sim$ 1800 of these `elementary observation intervals' (EOIs). In orbit
546, with its much lower count rate, each EOI consists of $\sim$ 100 sec
data, followed by the gap. Correspondingly, the Live Time of the detector
(column 6) is rather short for orbit 807 and highest for orbit 546.  Column
7 gives the average count rate recorded on the chip when no spatial nor
quality selection for the photons were made, while in column 8 we list the
screened, background subtracted source count rates for the total spectral
fit in the 0.6$-$10 keV band.
   
All data have been reprocessed using XMMSAS version 6.1. For the spectral
fits the latest response matrices were used, for which adjustments of
the effective areas around the Au edges were made as well as
corrections for the thick filter near the Al edge. We selected photons
with Pattern $\leq$ 4 (i.e. singles and doubles) and quality flag = 0.  
For the source data we selected the photons from rows 29 $\leq$ RAWX
$\leq$ 45 for our analysis, as background we used the photons from rows 2
$\leq$ RAWX $\leq$ 18 (for a description of the instrument and the
detector modes see Ehle et al. 2001). The energy range for all analyses
was taken to be 600 eV $\leq E \leq 10$ keV as below 600 eV the
single-double ratio is not constant and thus gives unreliable
results in the spectral analyses. The background count rates in the
whole energy band are generally less than $\sim$ 1.5\% of the source count
rates, and even at energies $\sim$ 10 keV they amount to $\simlt$ 10\% of
the source rates. A few short background flares and the data at the end
($\simgt$ 60 ksec)  of the observation of orbit 807 were discarded from
the spectral analysis presented in Sect. 4.
 
\subsection{Orbit integrated properties}

\begin{figure}
\psfig{figure=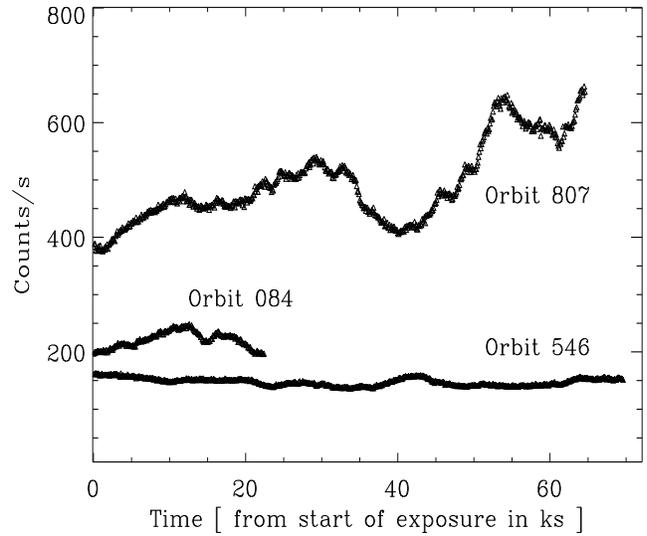,height=7.2truecm,width=8.5truecm,angle=0,%
 bbllx=80pt,bblly=370pt,bburx=535pt,bbury=700pt,clip=}
\caption[]{Background subtracted, 0.6$-$10~keV  PN  light curves of
Mrk 421 with time binning of 100 s.
Times are counted from the beginning of the actual exposure.
The curves are labeled by the orbit of the observation. }
\label{figure:light1}
\end{figure}

To demonstrate the magnitudes of the source variability we present in 
Fig. \ref{figure:light1} the  0.6$-$10~keV  100~s bin light curves 
for the three observations  with the above photon selection criteria.
The counting errors are smaller than the symbol sizes.
For comparison with previously published  light curves please note that 
the count rates are dominated by photons at low energies. For example,
extending the energy range down to 500 eV would increase the count rate
by about 15\%, going down to 200 eV nearly doubles the number 
of counts. The orbit 84 observation was performed with the
thick filter; correcting these data to the medium thick filter would result 
in a factor of typically $\simgt$ 1.2 higher count rates.

The light curves are characterized by long term trends lasting for time
scales comparable to or longer than the length of the observations and,
superimposed on this, of frequent flare like events of relatively regular
shape with characteristic time scales of the order of a few thousand
seconds. Interestingly, larger variations occur when the source is in a
higher state: during orbit 546 the maximum to minimum intensity variations
are $\simlt \pm $ 15\%, in orbit 84 of $\sim \pm $20\% and in orbit 807
even $\simgt \pm $ 40\%.
   
\subsection{Spectral analysis}

\begin{figure}
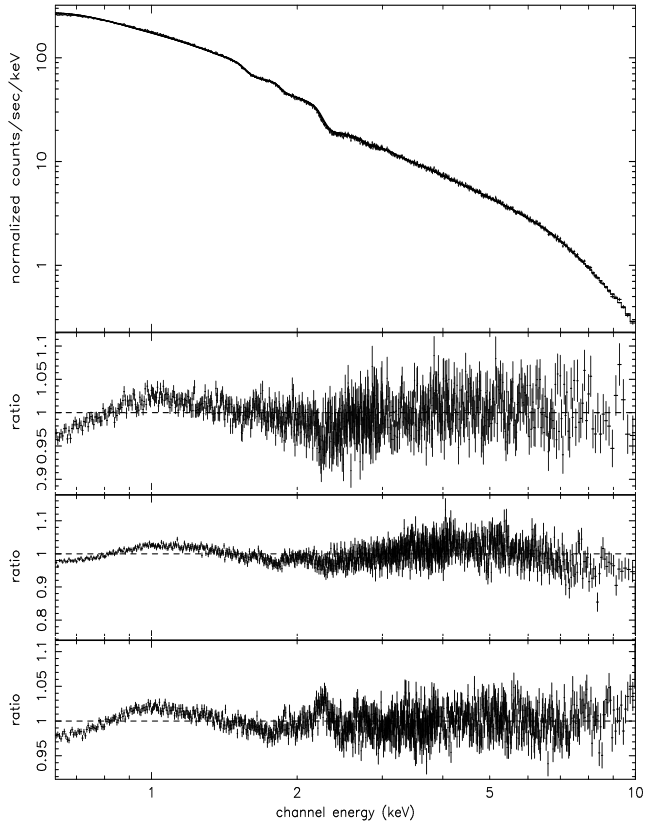

\psfig{figure=2767fg2a.ps,height=6.5truecm,width=8.5truecm,angle=-90,%
 bbllx=112pt,bblly=40pt,bburx=515pt,bbury=709pt,clip=}
\psfig{figure=2767fg2b.ps,height=1.9truecm,width=8.5truecm,angle=-90,%
 bbllx=383pt,bblly=40pt,bburx=514pt,bbury=709pt,clip=}
\psfig{figure=2767fg2c.ps,height=2.5truecm,width=8.5truecm,angle=-90,%
 bbllx=383pt,bblly=40pt,bburx=560pt,bbury=709pt,clip=}
\caption[]{\small Broken power law model fit to the 0.6$-$10~keV data
 during the three observations. The two top figures show the fit and the data to
 model ratio for orbit 84, the middle panel the ratio for orbit 546, 
 the bottom panel that for orbit 807. }
\label{figure:bknpo-tot}
\end{figure}

As mentioned above the spectral behavior of \M21 is rather complex and
with increasing data quality more sophisticated spectral models had to be
used for the fits. To show the integral spectral behavior of the source
during the three observations we present in Fig. \ref{figure:bknpo-tot}
the fits to the total data of each orbit to a broken power law.  The
plotted residuals are heavily rebinned. The deviations
between model and data are only of a few percent. The numerical
results are given in Table 2. The absorption has been fixed to the
galactic value; keeping N$_H$ as a free parameter results in slightly
lower values of $\chi^2_{\rm red}$ and slightly higher column densities.  
Interestingly, the slope changes at the break energy are rather small
which might indicate that we see the source near the Synchrotron peak of
its SED. Fitting a continuously curved model (Fossati \etal 2000b) to the
data results in fits of very similar $\chi^2_{\rm red}$, however the
residuals indicate that at higher energies the curvature of the spectrum
is over-estimated.

The $\chi^2_{\rm red}$ given in Table 2 show that the fits are hardly
acceptable which is to a great extent due to the extremely high photon
statistics in the soft energy band (for example, there are 13.2 million
photons in the orbit 807 spectrum).  There is still some uncertainty in
the calibration, visible around $\sim$ 2 keV (the Au edge and, possibly,
some Si feature at slightly lower energies) and a general convex curved
low energy part. Whether this slight extra low energy curvature is in the
range of the calibration uncertainties or a `source intrinsic' spectral
property cannot be answered yet. The residuals around 2 keV further
indicate that these structures are not strictly stationary during the
three observations taken over four years. 
They are likely not source specific but they could be small secular
changes of the detector characteristics over these years. 
(A gain-fit over a limited energy range around 2 keV reduces
the residuals strongly). Secondly, the spectral slopes vary with the
intensity variations causing some dispersion of the data with respect to
the average model. The time resolved fits of a broken
power law model to shorter data segments (see Sect. 4) mostly result in
$\chi^2_{\rm red} \sim 1.0$; thus a broken power law provides an adequate
description of these \M21 spectra. The general impression from the above
total fits is that the spectrum gets flatter with increasing flux and that
the break energy shifts towards higher values.

\begin{table}
\small
\tabcolsep1ex
\caption{\label{fits} Results for the broken power law fits 
assuming
galactic absorption (N$_{\mathrm H} = 1.66\times10^{20}$ cm$^{-2}$) }
\begin{tabular}{cccccc}
\noalign{\smallskip} \hline \noalign{\smallskip}
\multicolumn{1}{c}{Orbit} & \multicolumn{1}{c}{$\Gamma_{\rm soft} $} &
\multicolumn{1}{c}{E$_{\rm brk}$} & \multicolumn{1}{c}{$\Gamma_{\rm hard}$} &
\multicolumn{1}{c}{Norm} & \multicolumn{1}{c}{$\chi^2_{\rm red}$/dof} \\
\multicolumn{1}{c}{ } &\multicolumn{1}{c}{ } &
\multicolumn{1}{c}{(keV)} &\multicolumn{1}{c}{ } &
\multicolumn{1}{c}{ (1) } & \multicolumn{1}{c}{}\\
\noalign{\smallskip} \hline \noalign{\smallskip}
084 & 2.257$\pm$0.001 & 4.00$\pm$0.17 &2.354$\pm$0.009 & 0.182 & 1.22/1584\\
546 & 2.400$\pm$0.009 & 3.37$\pm$0.42 &2.620$\pm$0.006 & 0.108 & 1.69/1671\\
807 & 1.969$\pm$0.001 & 4.63$\pm$0.19 &2.027$\pm$0.006 & 0.316& 1.58/1900 \\
\noalign{\smallskip}\hline
\end{tabular}
\medskip   
   
 ~(1): Normalization at 1~keV in ph/keV/cm$^2$/s. \\
\end{table}

\section{Cross-Correlation Analysis}
  
In order to investigate the cross-links between the hard and soft band
light curves we estimated their Cross-Correlation Function (CCF), using
light curves with a 10-s bin size. If a data segment is followed
by the typical 21\,s long FIFO gap, this is filled with a running mean.
We originally chose to work with this small bin size because the signal is
rather strong and, secondly, previous cross-correlation analyses between
soft and hard band light curves have indicated that, if 
inter-band delays exist, they are very small (Brinkmann \etal 2003). 
The following temporal and spectral analyses show, however, that
noticeable spectral changes occur on longer time scales of a few hundred
seconds. Since the results from the previous section show that the energy
spectrum exhibits a significant spectral slope change at energies between
3 and 4 keV (see Table 2), we chose the count rates in the 0.6$-$2 keV and
4$-$10 keV bands as representative of the soft and hard band spectral
components of the source. This choice of energies further ensures that
there is no `spill-over' of flux between the two bands.

We calculated the sample Cross-Correlation Function, $CCF(k)$, as follows: 

\[
CCF(k)=\frac{\sum_{t}(x_{soft}(t) -\bar{x}_{soft})(x_{hard}(t+k)-\bar{x}_
{hard})}{N(k)(\sigma^{2}_{soft}\sigma^{2}_{hard})^{1/2}}, \]
 
\[ k=0,\pm \Delta t,\ldots,\pm (N-1)\Delta t. \]

The summation goes from $t=\Delta t$ to $(N-k)\Delta t$ for $k\geq 0$
($\Delta t = 10$ s, and $N$ is the total number of points in the light
curve)  and the sum is divided by the number of pairs included, i.e.
$N(k)$.  For negative lags $k < 0$ the summation has to be done over
x$_{\mathrm soft}(t +|k|)$ and x$_{\mathrm hard}(t)$. The variances in the
above equation are the source variances, i.e., after correction for the
experimental variance. Significant correlation at positive lags means that
the soft band variations are leading the hard band.

\begin{figure}
\psfig{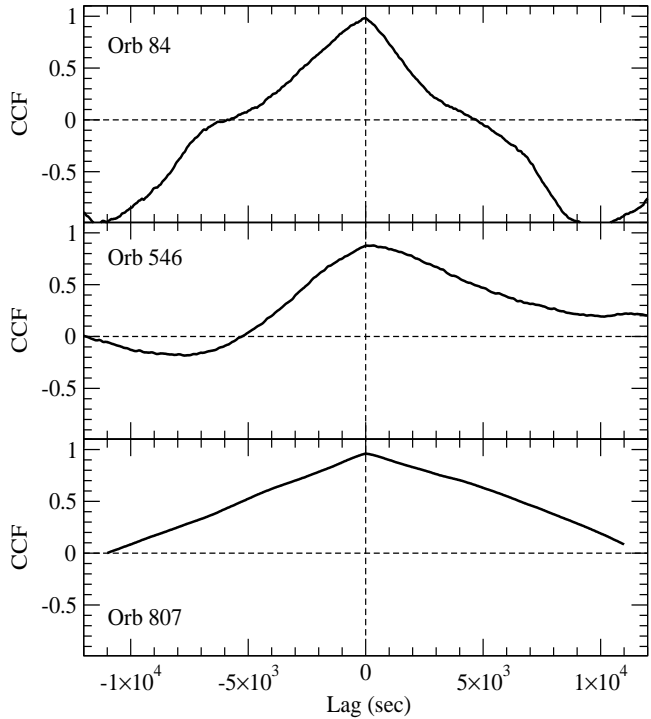}
\caption[]{\small CCF plots for the orbit 84 (top), orbit 546 (middle)
 and orbit 807 (bottom) observation.
The 10 s binned soft and hard band light curves for the total observations
were used; the error bars are smaller than the symbol sizes.
The curves are labeled by the orbit of the observation. }
\label{figure:ccftot}
\end{figure}

First, we computed the CCF (up to lags $\pm 3$ hrs) using the whole length
(i.e. total)  soft and hard band light curves. The resulting CCFs are
shown in Fig.  \ref{figure:ccftot} (the CCF for orbit 84 is similar to
Fig. 13 of Brinkmann \etal (2003), while Ravasio \etal (2004) present in
Fig. 9 the CCF of only the `flare' at t $\sim$ 42 ksec in orbit 546).
Although all three CCFs peak at zero lag, there are clearly definite
differences in shape between them.  The 
ranges for the values of the cross-correlation coefficients are different.
 Further, in
orbit 84 the auto-correlation function of the soft band is very similar to
the CCF; in the other two orbits the CCF is skewed with respect to the ACF
in the sense that the soft/hard bands are less correlated at negative lags,
but more correlated at positive lags $\simgt$ 1 hour.
 
This behavior, a peak at lag zero and a skewed CCF, has been observed many
times.  Ravasio et al. (2004) showed that such a shape can be be
reproduced by assuming flares that have the same form, a linear rise and
exponential decay, but with different amplitudes and rise/decay time
scales in the two energy bands. This explanation certainly works in the
case of well sampled isolated flares. Here, we do not focus on single
events but rather use the whole light curve to estimate the CCFs. The
observed light curves are characterized by many ``events'' rising or
decaying on different time scales, suggesting that more than one emission
region is in operation in the system. The emission mechanism in one of
them may operate on time scales equal to/longer than the duration of the
light curves. It causes the larger amplitude variations which happen in
phase in both energy bands, hence the zero lags when we integrate over the
whole length of the observation. On the other hand, emission from other
regions may operate on shorter time scales, causing changes of the
emission pattern during the observation on a ``characteristic time" scale
of just a few ksec. The physical meaning of this characteristic time
scale is not obvious. Observationally, in the present data it could
correspond to the duration of the flare-like events of relatively regular
shape in the light curves which is equal to a few thousand ksec. It is
possible that this time scale corresponds to the size of the individual
emission regions in the source. Weaker, shorter (i.e. on time scales of
seconds) variability might remain undetected in the measurement noise. On
the other hand, there are also longer time scales in the system
as, for example, the large scale variations during orbit 807 demonstrate.
We now investigate whether the correlation properties of the emission in
the two energy bands do evolve on periods comparable to this `few thousand
second' time scale.
 
Long integration intervals were required in the past to obtain an
acceptable signal to noise ratio with the limited photon statistics of
previous instruments. The largely increased sensitivity of \xmm allows the
analysis of much shorter data streams.  Brinkmann
\etal (2003) split the individual \M21 observations into two or
three sub-intervals with similar variability patterns and flux levels and
found differing CCFs for these sub-intervals.
 
Here we will employ a `sliding window' technique and calculate the CCFs
for shorter data intervals, which start at different times of the observed
light curve and range over a restricted time interval. Thus, the above
defined CCF(k) will then be replaced by a CCF(k; $\mathcal{T,L}$), i.e.,
the cross-correlation coefficient at a lag `k' is calculated for a data
stream with length $\mathcal{L}$ which starts at the time $\mathcal{T}$ in
the light curve.
 
\begin{figure}
\psfig{figure=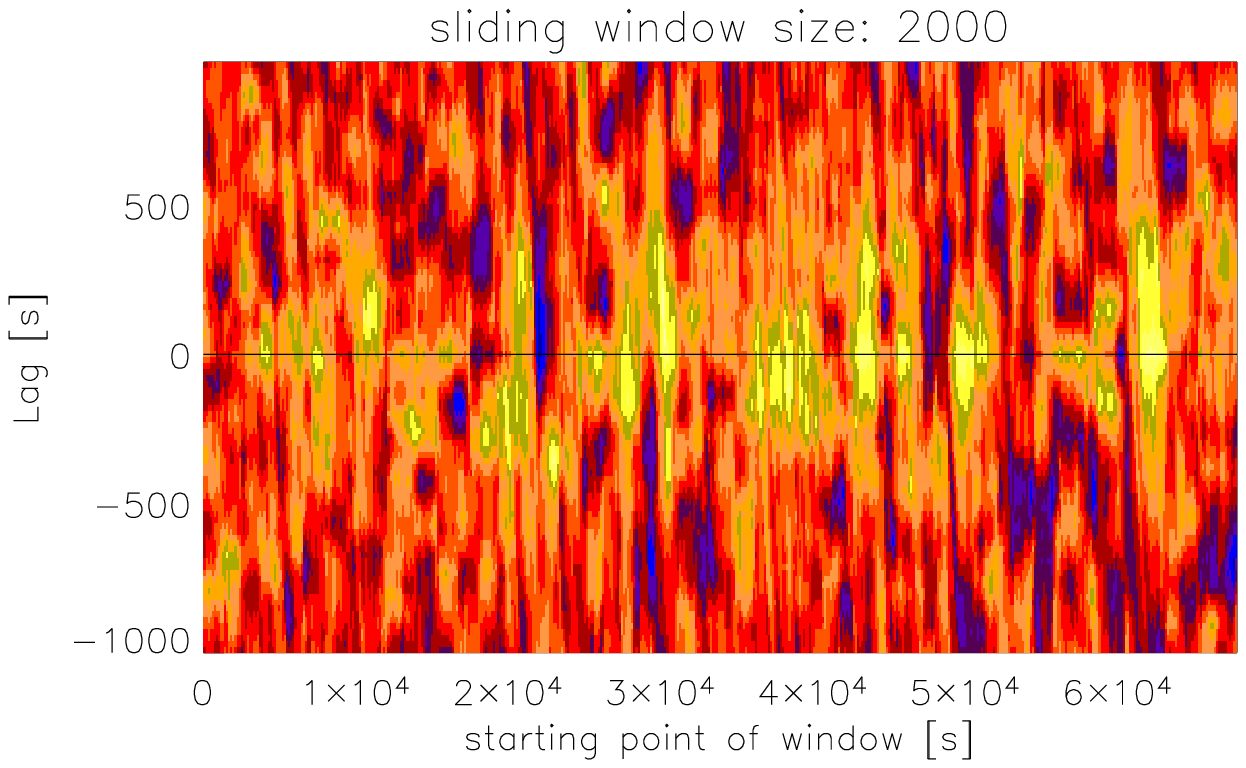,height=4.3truecm,width=8.5truecm,angle=0,%
 bbllx=52pt,bblly=351pt,bburx=415pt,bbury=540pt,clip=}
\psfig{figure=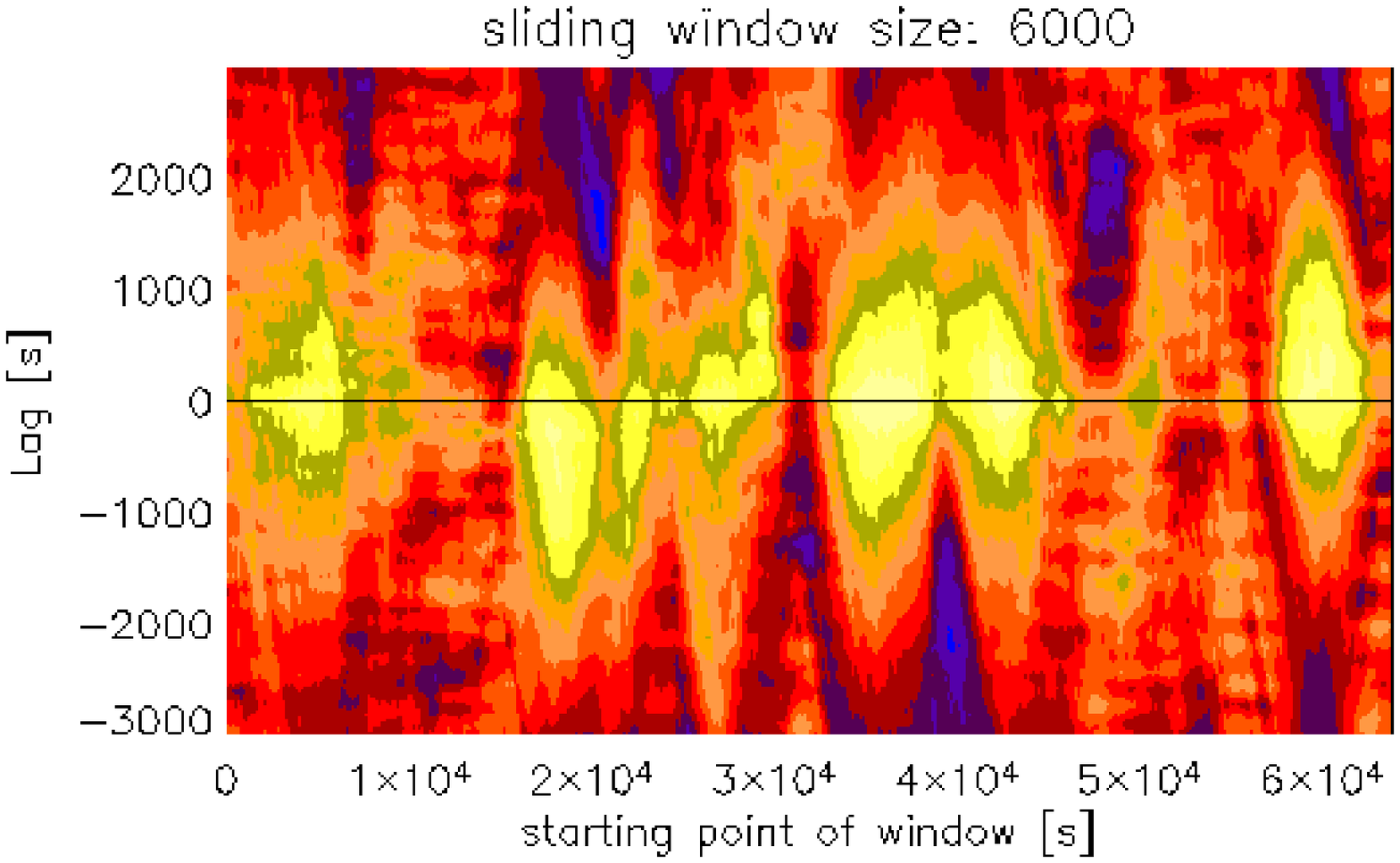,height=4.3truecm,width=8.5truecm,angle=0,%
 bbllx=35pt,bblly=261pt,bburx=558pt,bbury=540pt,clip=}
\psfig{figure=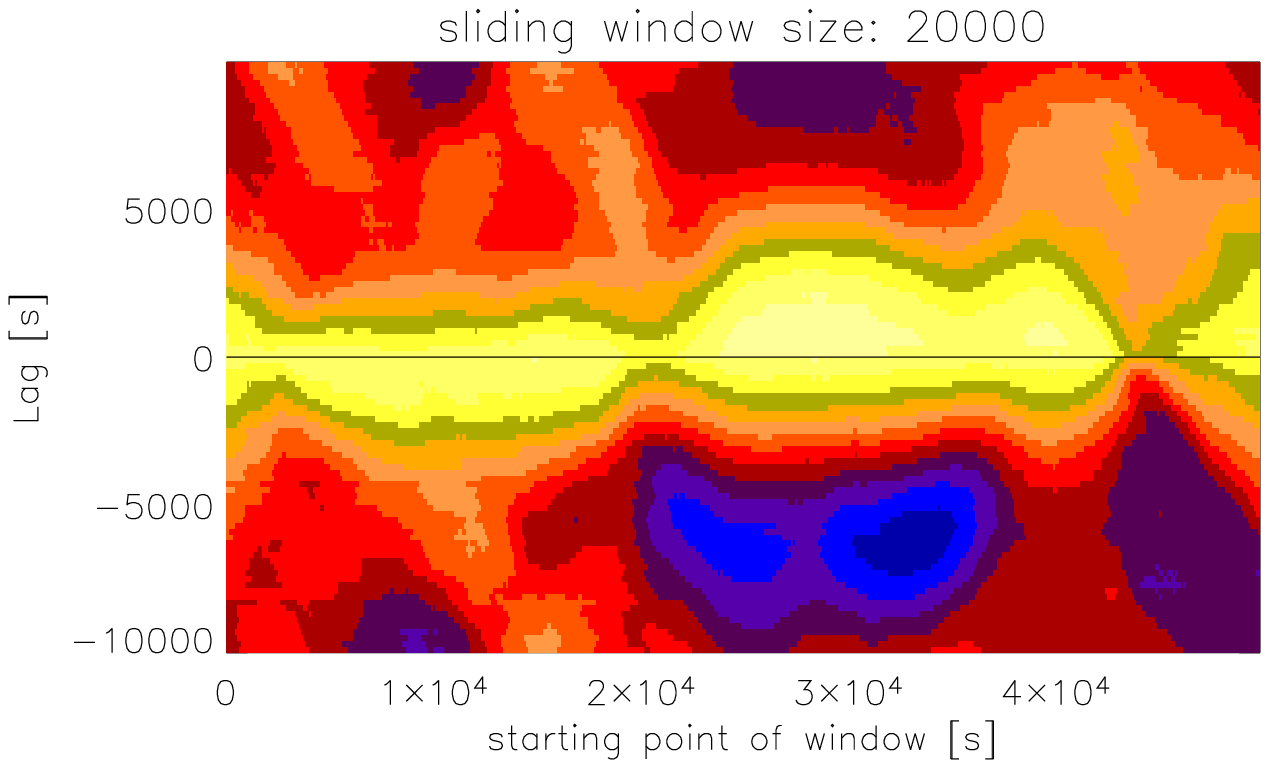,height=5.4truecm,width=8.5truecm,angle=0,%
 bbllx=52pt,bblly=328pt,bburx=415pt,bbury=540pt,clip=}
\psfig{figure=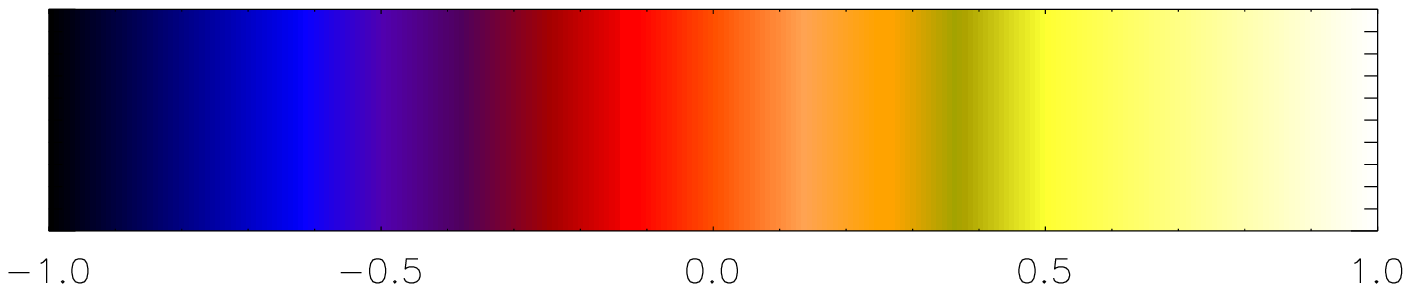,height=1.8truecm,width=8.5truecm,angle=0,%
 bbllx=11pt,bblly=1pt,bburx=485pt,bbury=85pt,clip=}
\caption[]{\small Sliding window CCFs for orbit 546. The 
 amplitude of the CCF is color coded, with lags plotted in the vertical direction.
 The top
figure was calculated with an individual data length of 2000~s, the middle 
 with 6000~s, and the bottom figure with 20000~s  long data streams. 
The bottom panel shows the color bar used for the numerical values of the CCFs
 for  the above figure and for Fig. \ref{figure:slidemax}. 
 }
\label{figure:slide1}
\end{figure}

In Fig. \ref{figure:slide1} we show a two dimensional representation of
these sliding window CCFs, based on light curves with 10\,s binnings. The
vertical scale represents the lags (similar to the x-axis of Fig. 3), the
color coding indicates the amplitude of the cross-correlation coefficient,
with the color code given in the lowest panel of the figure. Time is along
the x-axis. In the top figure the length of the individual data sets is
$\mathcal{L}$ = 2000~s (thus the longest part of the light curve is
covered), for the middle it is 6000~s and for the bottom figure 20000~s.
Clearly, in the first figure, CCF(k; $\mathcal{T,}$2000), the
cross-correlation coefficient is very noisy, the CCF is dominated by local
fluctuations and the data length $\mathcal{L}$ seems to be shorter than
the typical variability time scale.  The middle panel ($\mathcal{L}$ =
6000~s) shows varying lags of both signs with different amplitudes as well
as periods where there is hardly any inter-band correlation. This length
seems to be in the range of the variability time scale of the source and
it further demonstrates that the physical conditions of the emission
regions change on that time scale. In the bottom figure the integration
length of $\mathcal{L}$ = 20000~s over the data stream is obviously much
longer than the variability time scale.  Local variations of the sign and
amplitude of the lags are smoothed out and produce cross-correlations with
lags around zero, with an asymmetric amplitude distribution as found in
Fig. \ref{figure:ccftot}.
 
\begin{figure}
\psfig{figure=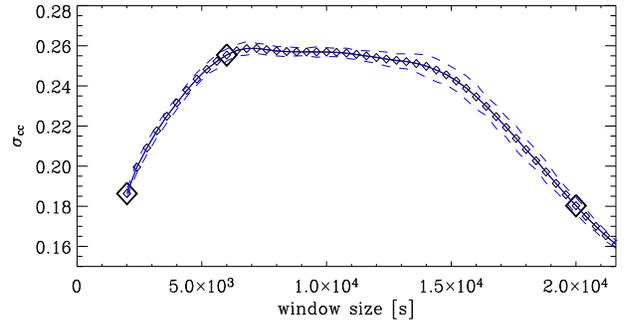,height=4.3truecm,width=8.5truecm,angle=0,%
 bbllx=6pt,bblly=6pt,bburx=415pt,bbury=196pt,clip=}
\caption[]{\small  Standard deviations $\sigma_{P(ccf)}$ of the 
 sliding window CCFs
 for orbit 546 for different data lengths  $\mathcal{L}$. The diamonds
are the  measured values, the three large symbols  
  indicate the window lengths used for the CCFs in 
 Fig. \ref{figure:slide1}. The solid line gives the mean and the dashed lines 
the standard deviation
of the re-sampled CCFs (see text).}
\label{figure:ccflang}
\end{figure}

The above discussion shows that there is an optimal window length
$\mathcal{L}$ which reveals best the time evolution of the CCF structure.
In order to determine this length we regard the values ('response') of the
'local' cross correlation functions, CCF(k), in all sliding windows for
lags -2000\,s $\leq$ k $\leq$ +2000\,s as a probability distribution
function (PDF), $P(ccf)$, of different response values ranging from -1 to
1. By determining the statistical properties of these PDFs as a function
of $\mathcal{L}$ we can identify the optimized size of $\mathcal{L}$. If
the sliding window is very small, only small correlations between the
bands can be identified, because the CCF is dominated by the random
fluctuations of the experimental noise.  This results in a relatively
narrow PDF peaking at $P(ccf) \approx 0$. If $\mathcal{L}$ is very large,
one is no longer sensitive to the temporal fluctuation of the CCF, which
may vary with appearing or disappearing flare events. Only an overall
`mean cross correlation' between the bands is measured. The resulting PDF
is also narrow but with a maximum at high values accounting for the
obvious fact that there is an overall correlation between the energy
bands.  However, if the sliding window has an intermediate `optimal'
length one is sensitive to the variations of the CCF, if present.  The
resulting $P(ccf)$ becomes broad with CCF-values covering a large part of
the whole possible range and no pronounced maximum.  To quantify this
anticipated behavior of the PDFs we calculated their standard deviation
$\sigma_{P(ccf)}$ and plot them as a function of the sliding window sizes,
$\mathcal{L}$ (Fig. \ref{figure:ccflang}).  The shape of the curve shows a
broad maximum with the highest value at a window length $\mathcal{L} \sim
$ 7.2 ksec.  At window sizes greater than $\sim 15$ ksec the standard
deviations gradually decrease.
 
To verify that the observed behavior is of physical nature and not caused
by statistical effects, namely less sampling for larger window sizes, we
performed the following validation, which is inspired by ideas from
bootstrapping (Efron 1993): For each window size we randomly selected as
many CCFs as we have for the largest window size, and use them as the
basis for the calculation of the standard deviation.  This procedure
enables us to determine a kind of bootstrap-error of the calculated
$\sigma_{P(ccf)}$ by repeating the re-sampling procedure many times and
calculating statistical means and standard deviations of the measured
quantity sigma. The results are shown in Fig. \ref{figure:ccflang} as the
three lines (mean and standard deviations).  It is obvious that we obtain
the same results as previously, when the number of windows changed with
$\mathcal{L}$. 
 
Therefore, the highest variations in the CCF(k; $ \mathcal{T,L} $)  at a
window length $\mathcal{L} \sim $ 7.2 ksec truly indicate that this value
matches the typical time scale in the light curve  and that cross
correlation functions calculated over the total light curve (i.e. $\gg $
20 ksec) provide only little information.
 
\begin{figure}
\psfig{figure=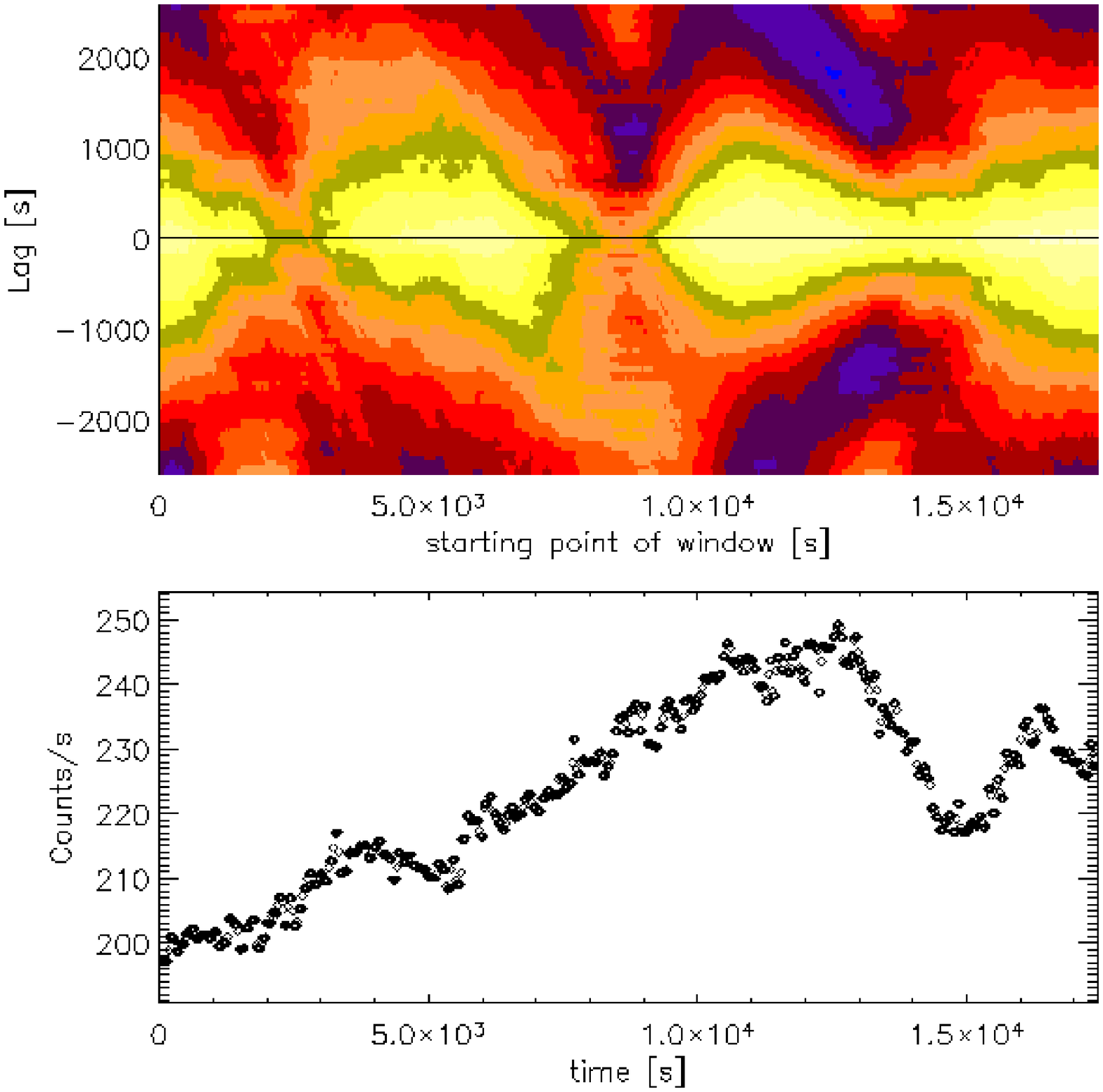,height=7.0truecm,width=8.5truecm,angle=0,%
 bbllx=18pt,bblly=-2pt,bburx=605pt,bbury=576pt,clip=}
\psfig{figure=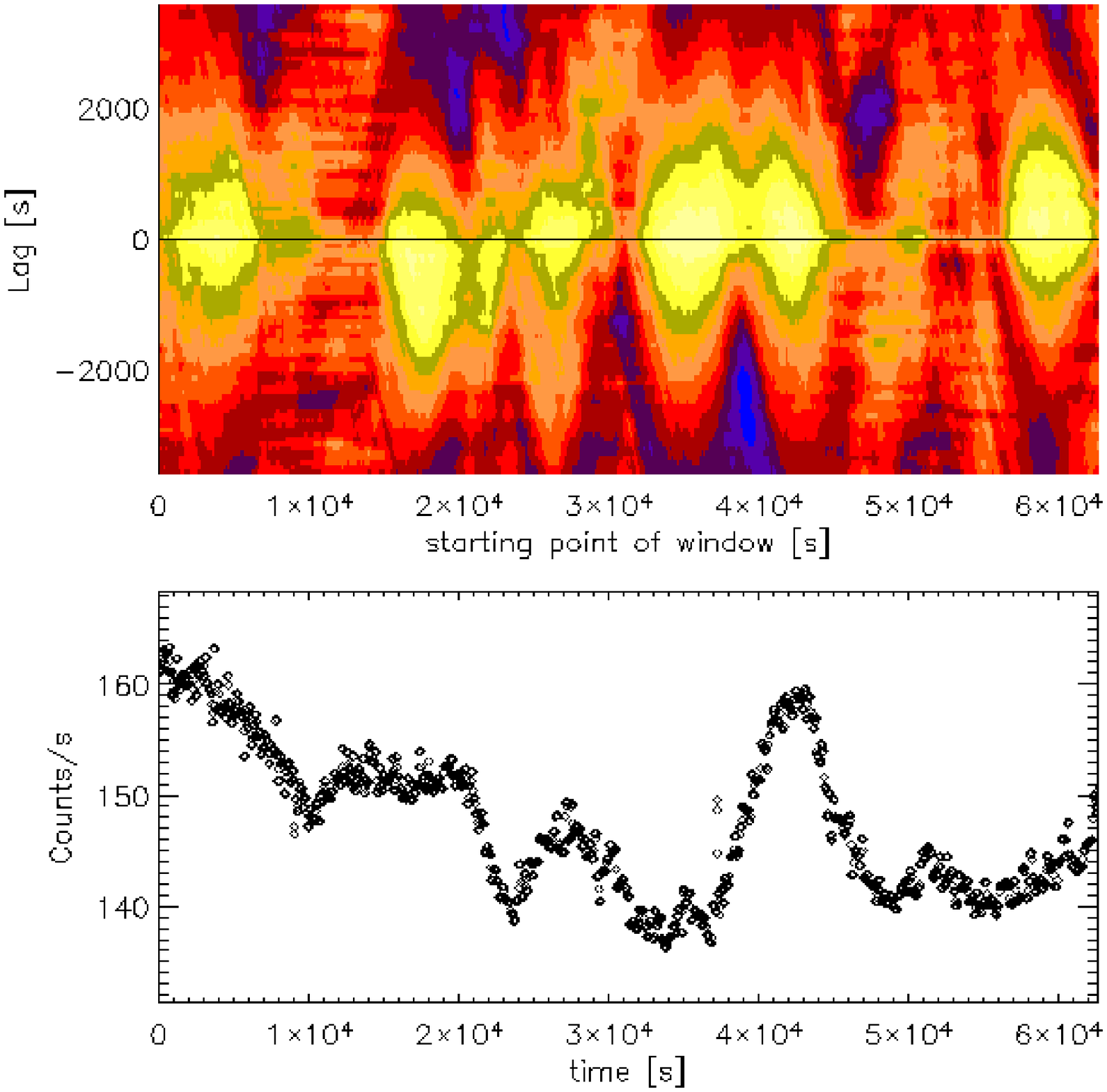,height=7.0truecm,width=8.5truecm,angle=0,%
 bbllx=1pt,bblly=106pt,bburx=589pt,bbury=685pt,clip=}
\psfig{figure=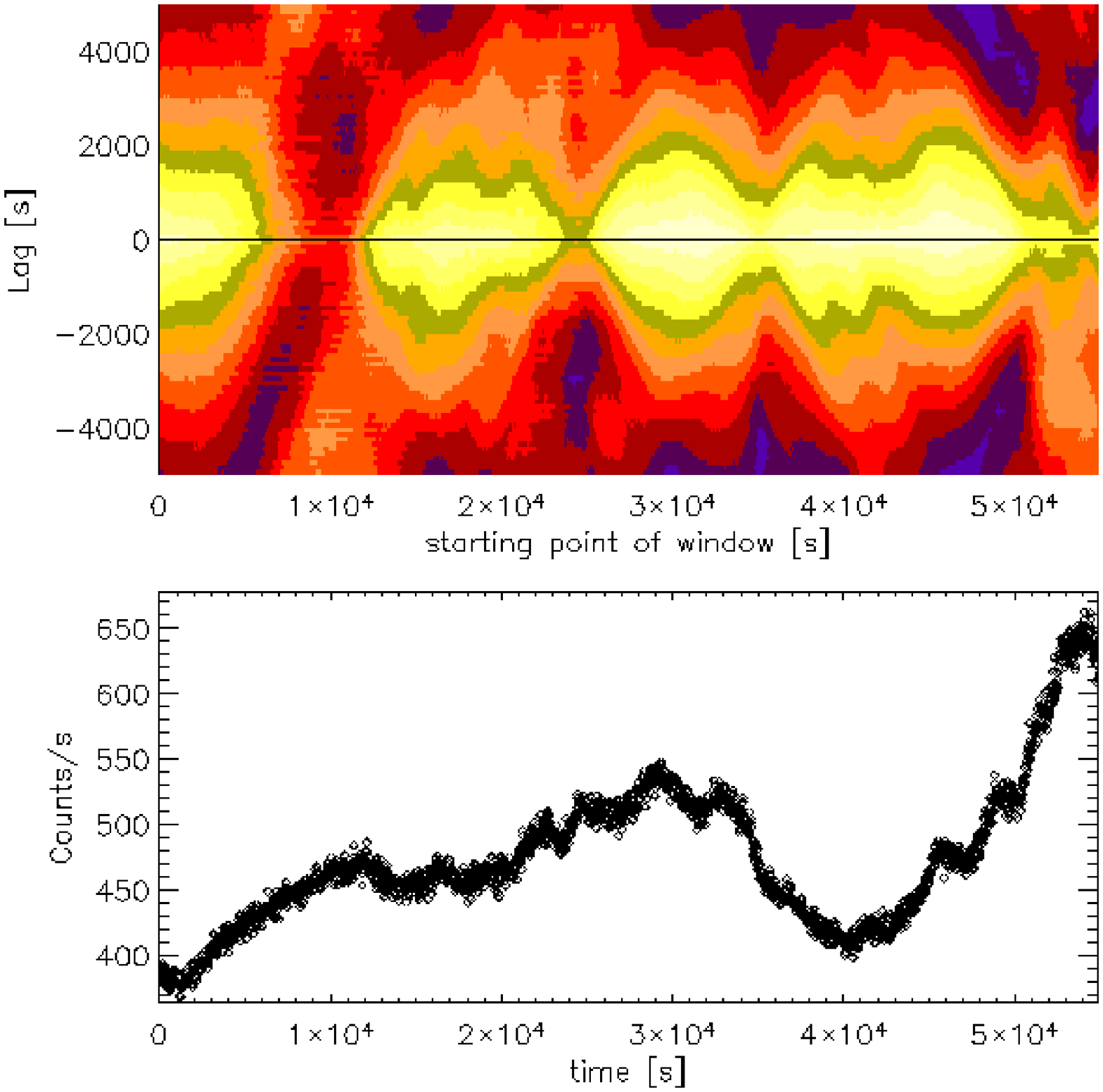,height=7.0truecm,width=8.5truecm,angle=0,%
 bbllx=6pt,bblly=95pt,bburx=593pt,bbury=677pt,clip=}
\caption[]{\small Sliding window CCFs for orbits 084 (top), orbit 546 (middle)
 and orbit 807 (bottom) calculated with the optimized sliding window lengths (see
 text). The amplitude of the CCF is color coded, with lags plotted in the  
 vertical direction. Plotted below the individual CCFs  are the corresponding
 light curves in the total (0.6$-$10 keV) energy band. The x-axis is the time 
 from the beginning of the observation  which, for the CCFs, corresponds to the
 start time of the data window for the calculations. 
 }
\label{figure:slidemax}
\end{figure}

For the two other orbits a similar behavior was found with slightly
different values for the maximal values (orbit 84: peak at 5.2 ksec;  
orbit 807: peak at 10 ksec). In Fig. \ref{figure:slidemax} we show the
CCFs for all three orbits calculated with these three optimized window
sizes, for orbits 084 (top), orbit 546 (middle) and orbit 807. 
Again the amplitude of the CCF is color coded, with lags plotted in the 
vertical direction.  Below the individual CCFs we show the corresponding light
curves in the total (0.6$-$10 keV) energy band. The x-axis is the time
from the beginning of the observation which, for the CCFs, corresponds to
the start time of the CCF calculation. The light curves end
at the time given by the difference of the observing time and the
length of the sliding window.
  
The plot clearly demonstrates the complex behavior of the energy resolved light
curves and the considerable uncertainties interpreting them.
Let us note some peculiarities which are hard to accommodate in
`standard' scenarios of the emission from the BL Lac jets:
\par \noindent {\bf Orbit 084:} 
Generally, there are always nearly zero lags between the two
energy bands. The correlation coefficient tends to zero from times
t $\sim$ 2$-$3\,ksec, at the small bump in the light curve, and at around 
8$-$9.5\,ksec, at the continuously increasing part of the light curve.
As the data length $\mathcal{L}$ of orbit 084 is taken to be 5.2 ksec,
the CCF at $\sim$ 2\,ksec, where it changes from highly correlated to weakly
correlated, is calculated from data between 2$-$7.2\,ksec; that where it 
changes back between $\sim$ 3$-$8.2\,ksec. Therefore, the data segments that
can cause those changes are either between 2$-$3\,ksec or between 7.2$-$8.2\,ksec,
i.e., at the beginning or at the end of the data sets. The light curve does not
provide a clear answer to this question; however, the time resolved spectroscopy 
(see next section, Fig. \ref{figure:084gam})  clearly points to a short hard
flare centered at t $\sim$ 3.5\,ksec. 
The  final parts of the CCF at times $\simgt 10^4$\,s, which cover the 
intensity drop by about
25\% at t $\simgt$ 13\,ksec and the subsequent recovery of the count rate
indicate that these flux variations have occurred at nearly zero lags.
\par \noindent {\bf Orbit 546:}
A small positive lag at the beginning, followed by zero lags 
during the following plateau phase. The intensity drop at t = 20\,ksec results in
a strong negative lag, while the increase and the subsequent decrease at 
 t $\simgt$ 38$-45$ \,ksec results in positive lags. The final parts of
 this decrease and the small substructure at $\sim$ 51\,ksec seem to have 
 occurred nearly uncorrelated at the two energy bands. Note that the absolute
variability amplitude during this observation is relatively small.
\par \noindent {\bf Orbit 807:}
The orbit where the source is brightest and exhibits the largest 
variability. Correlated at near-zero lags at the beginning and then, around the
turning point of the flux increase, an extended period of very low correlation.
The large variability region at t $\simgt$ 30\,ksec seems to have occurred in
both energy bands without strong time lags. However,  the broadness of the
 correlation function at these later periods could indicate that the time scales
 of the smaller intensity fluctuations are not resolved.
 
These investigations lead us to the conclusion that, quite generally,
 an analysis of light curves
in different energy bands might be inconclusive; in particular, when rapid and
strong intensity variations cannot be resolved temporarily, for example, due to
a low signal to noise ratio. 
The above CCF results demonstrate the complexity of the observed variations of
the light curves in different energy bands but do not provide  further 
information on the spectral behavior.
An intensity decrease/increase is once accompanied by soft/hard lags (respectively)
and once occurs nearly achromatically. A more detailed understanding of the
variability behavior of the source can be achieved with a study of the
variations of the spectral slopes on time scales as short as possible, which we
present below.

\section{Time-resolved spectroscopy}

The observed count rates in the PN camera are high enough to
study the energy spectrum of \M21 over short periods of time in
order to investigate in detail its spectral variability
properties. To this end, we extracted energy spectra over the
short observation intervals between the gaps that are
introduced when the detector falls into counting mode (see
Sect. 2 above). The typical duration of these intervals are
$\sim$ 100\,s, 80\,s and 20\,s for the observations of orbit
84, 546 and 807, respectively.  To avoid insufficient signal
for extremely short EOIs we had to add in some cases subsequent
data intervals to get the abovementioned `typical' data
stretches. Thus, for orbit 84 we fitted 149 individual data
segments, 476 for orbit 546 and 456 for orbit 807.  The number
of photons for the individual fits in the $0.6-10$ keV band
were between 15000 and 30000, which is large enough to perform
a meaningful model fitting to each spectrum individually.
We used the same response matrices as for the total energy
spectrum (see Sect. 2.2) and again fitted a broken power law
model.
  
As a first test of the physical relevance of our results we
investigated the best fit normalization values. We found an
excellent correlation between these values and the observed
count rate in the 0.6$-$10 keV energy band indicating that
the observed flux variations are mainly caused by variations in
the power law normalization (as opposed to spectral slope
variations).

In most cases, the best fitting break energy value turned out
to be in the range between $1.5-5$ keV, i.e. similar to the
values we found for the fits to the total energy spectra
of the source. In some cases though, the full band energy
spectrum could be quite well fitted by a single power law, and
thus either the hard or soft spectral slopes ($\Gamma_{hard}$
or $\Gamma_{soft}$, respectively) were determined with
significant uncertainty. Furthermore, in most cases the best
fitting $\Gamma_{hard}$ and $\Gamma_{soft}$ values are not
different by more than $\sim 1-2 \sigma$. This is due to the
fact that the uncertainty associated with the hard band
spectral slope is often quite large which is a direct result of
the fact that most of the photons in the individual spectra are
in the soft energy band. However, as our aim is to
investigate the intrinsic temporal variations of the spectral
``shape" of the source, the somewhat larger uncertainties of
the individual best fitting slope values are not so important.
What {\it is} important is to investigate whether the best
fitting values are {\it systematically} different between
successive data stretches, and if so, in what way.

\begin{figure}
\psfig{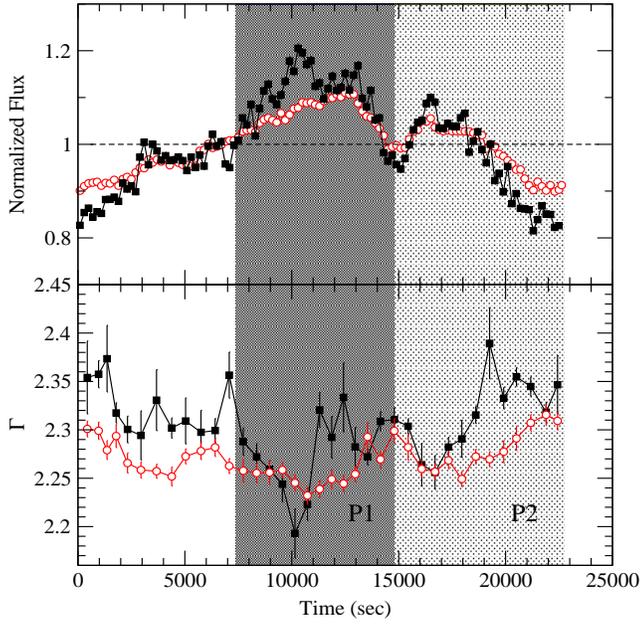}
\caption[]{Background subtracted, 0.6$-$2 and 4$-$10 keV PN light curves
of Mrk 421 in orbit 84 (open circles and filled squares, respectively), in
bins of size 200 sec (top panel). The light curves are normalized to their
mean. The corresponding $\Gamma_{\mathrm hard}$ and $\Gamma_{\mathrm soft}$ 
curves (filled squares and open circles, respectively) 
are plotted in the bottom panel (binned as described in the text). The
shaded boxes identify those parts of the observation where the $\Gamma$ vs
source flux relation is studied in more detail (see Sect. 4.1).}
\label{figure:084gam}
\end{figure}

\begin{figure}
\psfig{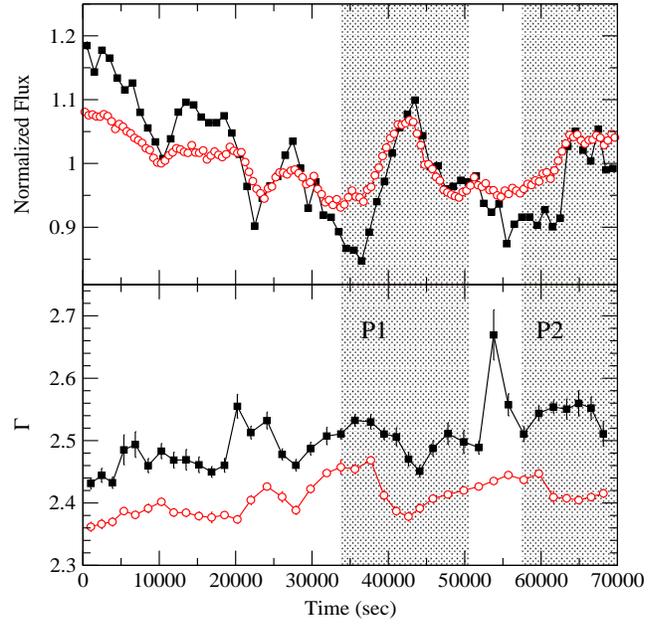}
\caption[]{Same as Fig.~\ref{figure:084gam}, but for the observation
during orbit 546. The hard band and soft band count rate light curves are
in bins of size 1000 s and 500 s, respectively.}
\label{figure:546gam}
\end{figure}

\begin{figure}
\psfig{figure=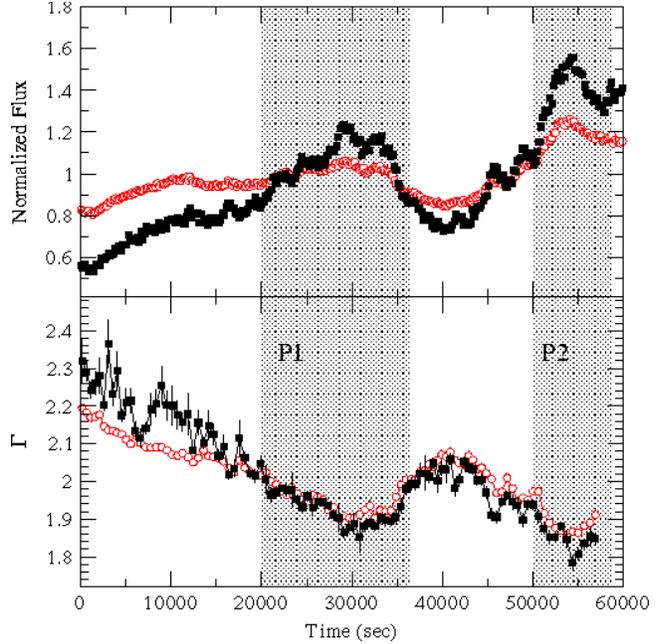,height=8.5truecm,width=8.5truecm,angle=0,%
 bbllx=49pt,bblly=120pt,bburx=562pt,bbury=645pt,clip=}
\caption[]{Same as Fig.~\ref{figure:084gam}, but for the observation
during orbit 807.The count rate light curves are in bins of size 40 s.}
\label{figure:807gam}
\end{figure}

Due to their large uncertainties and the existence of a few
outliers, the light curves of the best fitting hard and soft
slope values are noisy. For that reason we binned the
respective light curves by calculating the weighted mean of 4
consecutive slope estimates in the case of orbits 84 and 807
observations and of 12 values for orbit 546. As a result, we
are able to study the spectral slope variations of the source
on time scales as short as $\sim$ 500\,s and $\sim$ 700\,s in
the case of orbits 84 and 807, and $\sim$ 180\,s for orbit 546.
The bottom panels in Figs. \ref{figure:084gam}$-$\ref{figure:807gam} 
show plots of the
binned $\Gamma_{\mathrm hard,soft}$ light curves. The soft and
hard band count rate light curves are plotted in the top panels
of the same figures.

These figures show clearly that apart from the strong, fast
flux variations, \M21 also exhibits significant spectral
variations on time scales as short as $\sim 500-1000$ sec. To
the best of our knowledge, this is the first time that 
these quantitative spectral variations on time scales of a few
hundred seconds have been studied along with the flux variations.
We observe significant slope variations with a max-to-min
variability amplitude of the order of $\sim 7-10$\% and $\sim
4-6$\% in the hard and soft slope light curves, respectively,
for orbits 84 and 546. Apart from the largest amplitude flux
variations, orbit 807, when the source is at its brightest
state, shows the largest amplitude spectral variations as well:
the max-to-min amplitude is $\sim 30$\% and $\sim 20$\% for the
$\Gamma_{hard}$ and $\Gamma_{soft}$ light curves, respectively.
In all cases, the hard band slope variations are of larger
amplitude when compared with the soft band slope variations.
Furthermore, both of them are of smaller amplitude than the
amplitude of the flux variations in the respective bands.

Apart from the significant, fast spectral variations, the
bottom panels of  Figs.~\ref{figure:084gam}$-$
\ref{figure:807gam} show that, most of
the time, the hard and soft band slopes are {\it not} the same.
This result justifies a posteriori the use of the broken power
law model fit to the individual, time resolved spectra of
the source. Although the uncertainties of the 
slope values from the fits to {\it each}
individual spectrum are not negligible the 
availability of many spectra, {\it and} the use of a broken
power law model, reveals clearly that the soft and hard 
band slopes differ in a {\it systematic} way. In most cases,
the slope of the photon spectrum at hard energies is steeper
than the soft band slope. There are cases though  when the
photon spectrum is well described by a single slope power law
(for example the period between 20$-$30 ksec after the start
of the observation in orbit 807) and even cases when the hard
band slope is {\it flatter} than the soft band slope (for
example the period after the first 30 ksec in orbit 807).

The light curves of the fitted break energies show
significant variations which are neither correlated with the
best fitting slopes nor with the sources' flux states. This may
well be representative of the intrinsic behavior of the source.
On the other hand, as discussed in Sect. 2.2, the X-ray
spectrum of \M21 may show a ``continuous"  curvature which the
broken power law model can only approximate to some level.
Thus, the break energy may not correspond to an intrinsic
characteristic of the source, but may simply ``divide" the
spectrum into the two regions that show the largest difference
in ``curvature". For this reason, we do not study the break
energy variations hereafter. Instead, we concentrate on the
study of the $\Gamma_{hard}$ and $\Gamma_{soft}$ variations,
having in mind that these slopes simply give a measure of the
spectral curvature towards the soft and hard end of the X-ray
spectrum.

\subsection{The relation between spectral and flux variations}

``Hardness (or softness) ratio vs count rate" plots have
been used in the past mainly to study the correlation between
the flux and spectral variations in this source and other BL
Lacs (see for example Brinkmann et al. (2003) and Ravasio et
al. (2004) for a recent use of these plots with \xmm\ data). In
some cases, the evolution of the spectral slope itself as a
function of the source flux has also been investigated. For
example, Takahashi \etal (1996) studied the spectral slope
variations of the source based on a day-long ASCA observation,
Fossati \etal (2000b) used the best continuously curved
model fits to study the spectral evolution on time
scales of $\sim 10$ ksec, based on BeppoSAX observations,
while Ravasio \etal (2004) have used simple power law
fits to study the spectral variations during a few
particular periods, chosen from two \xmm\ light curves.

We believe that the analysis presented in the
previous section provides a further improvement in these
studies. Due to the high sensitivity of \xmm\ we
can probe spectral variations on time scales as short as $\sim
1$ ksec, based on model fitting results rather than on hardness
ratios. By fitting a broken power law we are
able to investigate separately the relation between the slopes
of the soft and hard bands of the spectrum with the count
rates in the respective energy bands. Also, we will study the
cross-correlation between the soft and hard band slope light
curves. 

Figs.~\ref{figure:084gam}$-$\ref{figure:807gam} show that 
the hard and the soft band slope
variations are correlated to a large degree. In general, both
parts of the spectrum steepen or flatten at the same time,
although not always at the same rate. There are cases when the
hard and soft band variations are somehow ``disconnected". For
example, in the period around 7 ksec, between $11-13$ ksec 
and at 18 ksec during orbit 84, we observe fast, large amplitude
$\Gamma_{hard}$ variations, which are absent in the
$\Gamma_{soft}$ time series. Similar events are observed in
orbit 546, and  at $7-11$ ksec in orbit 807. Not surprising,
during the same periods, the peak in the CCF between the soft
and hard band light curves is decreased (Sect. 3).

\begin{figure}
\psfig{figure=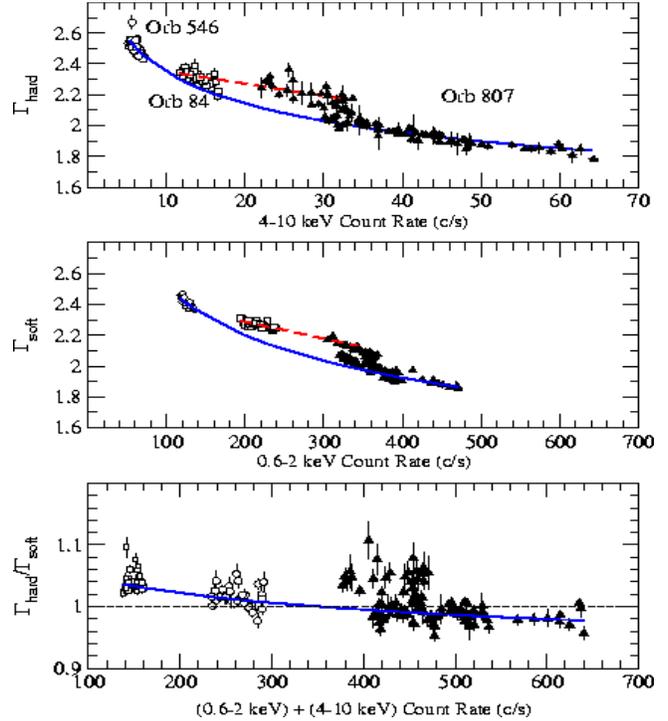,height=9.5truecm,width=8.5truecm,angle=0,%
 bbllx=50pt,bblly=130pt,bburx=476pt,bbury=672pt,clip=}
\caption[]{The $\Gamma_{hard}$ vs $4-10$ keV count rate and the
$\Gamma_{soft}$ vs $0.6-2$ keV relations (top and middle panels,
respectively). The bottom panel shows the ratio
$\Gamma_{hard}/\Gamma_{soft}$ plotted as a function of the $0.6-10$ keV
count rate. Data points from Figs.~\ref{figure:084gam},
\ref{figure:546gam}, and \ref{figure:807gam} are shown with open circles,
open squares and filled triangles, respectively. The solid curves describe
broad trends as discussed in the text. 
For the
orbit 84 observation, which was performed with a thick filter, the
$0.6-2$, $4-10$ and $0.6-10$ keV count rates are multiplied by factors of
1.22, 1.01, and 1.19, respectively. These are the count rate ratios
expected for the thick and medium thick filters in the case of a power law 
spectrum and N$_H$ similar to that of \M21.}
\label{figure:figgcr}
\end{figure}

In order to investigate this anti-correlation further, we plot
in Fig.~\ref{figure:figgcr} the slopes $\Gamma_{\mathrm hard}$
and $\Gamma_{\mathrm soft}$ as a function of the soft and hard
band count rates, using all the data shown in
Figs.~\ref{figure:084gam}$-$\ref{figure:807gam}.
 The top and middle panel in this figure
show clearly the strong anti-correlation between the spectral
and flux variations: both the hard and soft band spectra become
harder as the flux decreases. This is a well known behavior for
this source and other BL Lacs (c.f. Pian 2002, and references
therein; Brinkmann et al. 2003).

The flux - spectral slope relation is obviously not linear.
Both the hard and the soft band slopes appear to ``saturate" to
the same limiting value of $\Gamma\sim 1.8$ at the highest flux
levels. This has already been observed in the past, see e.g.
Ravasio \etal\ (2004). However, the large number of data points
in Fig.~\ref{figure:figgcr} suggest a more complicated
picture. For example, although most points in the spectral
slope vs flux plots do follow a power law like relation (shown
with the solid lines in the two upper plots), 
 in other cases a linear relation
appears to describe the data better. It seems possible that
there are various well defined paths (i.e. ``spectral states")
in the "$\Gamma$ vs flux" plane that the source follows at
different times.

The bottom panel in Fig.~\ref{figure:figgcr} shows a plot of
the ratio $\Gamma_{hard}/\Gamma_{soft}$ versus the total (i.e.
$0.6-10$ keV) source count rate. We find that the 
slope ratio does not remain constant,
but rather decreases with increasing source flux. A power law
model (shown as a solid line in this panel) appears to be
broadly consistent with the decrease of the slope ratio as the
source flux increases, although not all points follow the same
trend.

\subsection{Temporal correlations of the spectral variations}

\begin{figure}
\psfig{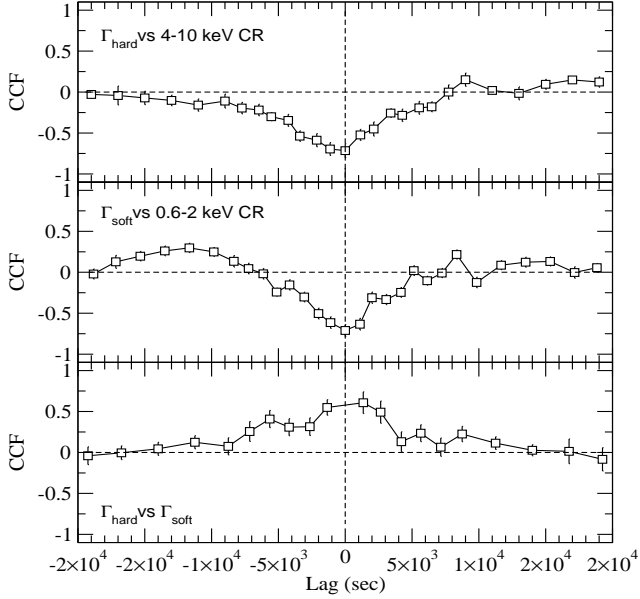}
\caption[]{The average cross-correlation function between $\Gamma_{hard}$
and $4-10$ keV count rate (top panel), the $\Gamma_{soft}$ and $0.6-2$ keV
count rate (middle panel), and between the two spectral slopes
$\Gamma_{hard}$ and $\Gamma_{soft}$ (bottom panel), using data from all
observations. The CCFs are defined such that any positive
delays would imply a delay of the spectral with respect to the flux
variations (in the upper two panels), and a delay of the $\Gamma_{hard}$
with respect to the $\Gamma_{soft}$ variations, in the bottom panel.}
\label{figure:gamlag}
\end{figure}

To investigate whether the spectral and flux
variations happen simultaneously, we computed the
cross-correlation function (CCF) between the spectral slope and
count rate light curves shown in 
Figs.~\ref{figure:084gam}$-$\ref{figure:807gam}. We used the
``Discrete Correlation Method" of Edelson \& Krolik (1988),
with a lag size of $\Delta k=800$ sec, 1000 sec, and 500 sec in
the cases of orbits 84, 546 and 807, respectively.  We also
computed the DCF between the $\Gamma_{hard}$ and
$\Gamma_{soft}$ light curves,  to examine the
simultaneity of the different band slope variations. Note that
in this case we did not estimate the CCF at lag zero, because
the slopes were obtained from model fitting to the same
spectrum and  any correlated, systematic biases in the
estimation of $\Gamma_{soft}$ and $\Gamma_{hard}$ can 
artificially increase the CCF value at this lag.

The cross correlation functions of all three observations look
very similar. For that reason, in order to increase the
signal-to-noise, we combined the three individual
$\Gamma_{hard}/\Gamma_{soft}$ vs flux and the $\Gamma_{hard}$
vs $\Gamma_{soft}$ CCFs into one file, respectively, and 
estimated average CCFs in each case.
The average $\Gamma_{hard}$ vs flux, $\Gamma_{soft}$ vs flux
and $\Gamma_{hard}$ vs $\Gamma_{soft}$ CCFs are shown in
Fig.~\ref{figure:gamlag}. The highest peak is at lag $\sim 0$
in all cases, suggesting that any delays between the slope and
flux variations, or between the soft and hard band slope
variations, should be shorter than $\sim 1000$ sec. The CCFs
look fairly symmetric around zero lag suggest that
both the hard and soft band spectra and flux variations happen
almost simultaneously, to the limit of the current temporal
resolution.

\begin{figure}
\psfig{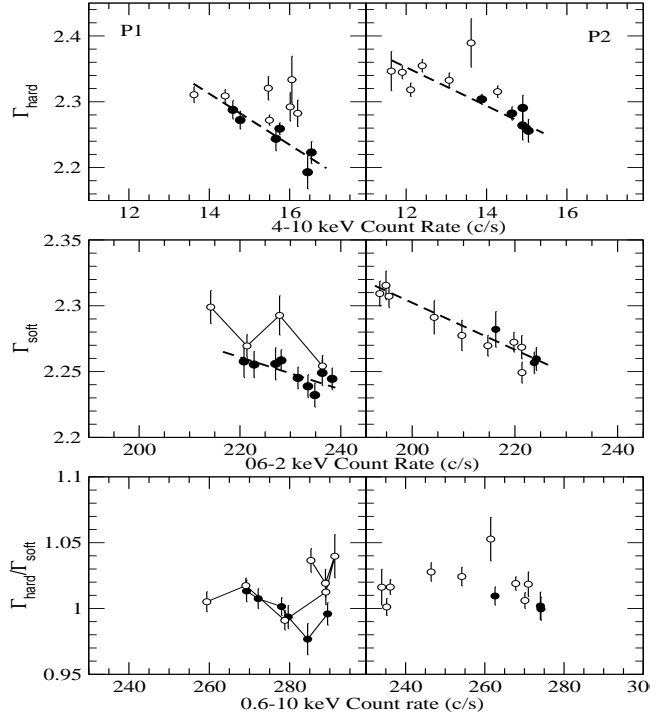}
\caption[]{Orbit 84: the $\Gamma_{hard}$ vs $4-10$ keV count rate and the
$\Gamma_{soft}$ vs $0.6-2$ keV relations (top and middle panels,
respectively) for the ``P1" and ``P2" parts (identified
by the shaded boxes in Fig.~\ref{figure:084gam}). The bottom panel shows
the ratio $\Gamma_{hard}/\Gamma_{soft}$ as a function of the
$0.6-10$ keV count rates. Filled circles show the data
 points during the flux rise and open circles during the decaying phase of each
event. The dashed line indicates the linear path that the rising (or the
decaying) phase points follow. For clarity, in cases where
the two phase points follow different paths, or define loop-like
structures, the respective points are connected with solid lines.}
\label{figure:84parts}
\end{figure}

\begin{figure}
\psfig{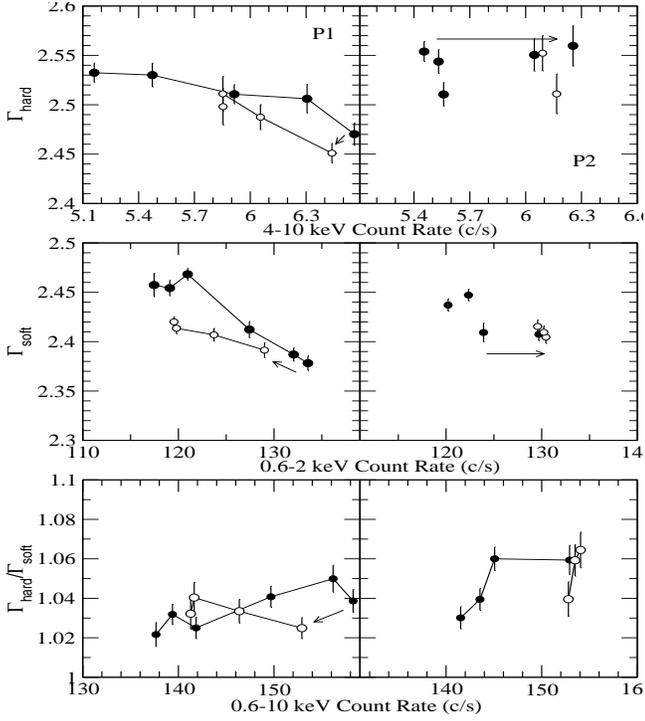}
\caption[]{Same as Fig.~\ref{figure:84parts}, but for the parts ``P1" and
``P2" in orbit 546.}
\label{figure:546parts}
\end{figure}

\begin{figure}
\psfig{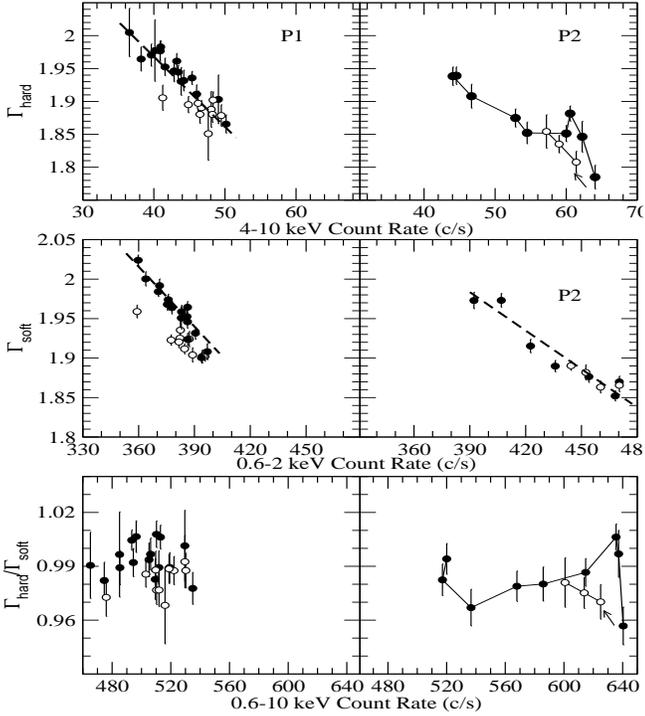}
\caption[]{Same as Fig.~\ref{figure:84parts}, but for the parts ``P1" and
``P2" in orbit 807.}
\label{figure:807parts}
\end{figure}

\subsection{The shortest time scales}

Although the results from Sect. 4.1 show that, in
general, the spectrum hardens as the source flux increases, the
$\Gamma$ vs flux behavior of the source is quite complicated
during individual "events". For that reason, we now focus our
study of the spectral variability behavior of the source during
the periods which are marked with shaded boxes in
Figs.~\ref{figure:084gam}$-$\ref{figure:807gam}. 
These correspond to well defined, single
``flare-like" events, where both the flux rising and decaying
parts have been adequately sampled (with the exception of the
P2 event in orbit 546, where it is not clear whether the flux
has started decreasing or not).
Figs.~\ref{figure:84parts}$-$\ref{figure:807parts} show the
$\Gamma_{hard}$ vs $4-10$ keV count rate (top), the
$\Gamma_{soft}$ vs $0.6-4$ keV (middle) and the
$\Gamma_{hard}/\Gamma_{soft}$ vs $0.6-10$ keV (bottom panels)
during these events. Filled and open circles correspond to data
points during the flux rise and decay phases, respectively.

The results from the analysis presented in this section are
complementary to those presented in the past (e.g. 
Takahashi \etal 1996, Fossati \etal 2000b, Ravasio
\etal 2004). However, they also provide some new insight
into the source's behavior. By a) studying all flaring events
that we could identify in the \xmm\ light curves, by b) using
spectral slope estimates in two energy bands 
on time scales as short
as $\sim 1$ ksec, and by c) combining spectral with timing
information (from Sect. 3) we demonstrate that the source
does {\it not} operate in the same way at all times.

Our results in the case of the P1 flare during orbit 546 are in
agreement with the results of Ravasio \etal\ (2004).
Although our model fitting results show clearly that the
the soft and hard band parts of the spectrum have different
slopes, both $\Gamma_{hard}$ and $\Gamma_{soft}$ follow a
clockwise loop pattern during the evolution of the flare. In
fact, even the $\Gamma_{hard}/\Gamma_{soft}$ ratio follows a
clockwise loop as the flare evolves. As the flux rises, both
the soft and hard band spectra flatten, but at different rates:
$\Gamma_{soft}$ flattens faster. 
When the flux decreases the spectra become steeper, 
but $\Gamma_{hard}$ steepens
faster than $\Gamma_{soft}$, and the slope ratio decreases for
a while. Finally, it increases again as $\Gamma_{hard}$ does not
vary appreciably any more while $\Gamma_{soft}$ still
increases. The CCF results suggest hard lags during the same
period (see Sect. 3), a result that is consistent with the
spectral evolution behavior of the source: variations are
propagated from the softer to the harder energy bands during
the initial flux rising phase, and then the hard band flux
decays faster than the soft band flux.

However, the source does {\it not} show the same behavior at
all times. For example, during the P1 flare in orbit 807 the
$\Gamma_{hard}$ variations follow the same path during the flux
rise and decay phase (shown with the dashed line in the top
left panel of Fig.~\ref{figure:807parts}). No loop patterns are
observed in the $\Gamma_{hard}/\Gamma_{soft}$ ratio either; 
the slope ratio remains roughly
constant, implying that the hard and soft band spectra evolve
at the same rate. The CCF results imply no delays
between the variations observed in the two bands. A similar
behavior is also observed during the P2 flare in orbit 84. Both
$\Gamma_{hard}$ and $\Gamma_{soft}$ evolve in the same way
during the rising and decaying phases (dashed lines in the top
and middle right panels in Fig.~\ref{figure:84parts}).
Consequently, the slope ratio remains roughly constant as the
flare evolves. In agreement with the observed spectral
variability behavior, the CCF analysis shows no delays between
the hard and soft band during this event.

The P1 flare in orbit 84 is a case where obvious variations in
$\Gamma_{hard}$ are not evident in the $\Gamma_{soft}$ time
series. The dashed line in the top left panel in
Fig.~\ref{figure:84parts} shows the $\Gamma_{hard}$ variations
as the flux increases (filled circles). Some of the
$\Gamma_{hard}$ values during the flux decay phase do follow
the same path, but most of the $\Gamma_{hard}$ values during
the prolonged "flare-plateau" phase between $11-13$ ksec
(Fig.~\ref{figure:084gam}) are clearly not consistent with it.
Furthermore, the $\Gamma_{soft}$ values during flux decay lie
systematically above the $\Gamma_{soft}$ values in the rising
phase suggesting the presence of an anti-clockwise loop
pattern. As a result of these two effects, the slope ratio also
follows a loop pattern, which evolves in an anti-clockwise
direction. The P2 flare in orbit 807 is another case of a flare
where the $\Gamma_{hard}$ and $\Gamma_{soft}$ variations do
not show the same behavior: the former follows a clockwise
loop pattern (top right panel in Fig.~\ref{figure:807parts}),
while the $\Gamma_{soft}$ evolution follows the same linear
path in both the rising and decaying flare phases.

That obvious case where the hard and soft band slopes are
almost entirely ``disconnected" is the P2 event in orbit 546.
Initially, $\Gamma_{hard}$ steepens ($\Delta\Gamma\sim -0.05$)
and $\Gamma_{soft}$ flattens ($\Delta\Gamma\sim 0.05$) although
the flux does not change appreciably. Then, both slopes remain
constant during a $\sim 15\%$ and $\sim 8\%$, rapid hard and
soft flux increase, respectively. Towards the end of the
observation, $\Gamma_{soft}$ remains constant, while
$\Gamma_{hard}$ flattens slightly (right panels
in Fig.~\ref{figure:546parts}). However, the slope ratio
changes imply that these apparently different variations are
in fact ``connected" in some way. A well defined loop pattern
appears in the slope ratio evolution plot which implies that
the variations are propagated from the soft to the hard band. 
This result is also supported by the strong, hard lags that
are observed in the same period (see Sect. 3).
 
\section{Discussion}
 
In this work we presented a spectral and timing analysis of the
three calibration phase \xmm observations of Mrk~421 performed
in timing mode. The analysis strongly relied on that mode as,
for example, the enormously high count rates of $\sim$ 800
cts/s in orbit 807 would have lead to a substantial photon pile
up - even in the extremely short frame time of $\sim$ 5.7\,msec
of the PN small window mode and thus to a large reduction of
photons available for the analysis.
 
Our average results confirm and, in a way, expand the findings
of previous \xmm investigations of \M21 (Brinkmann \etal 2001,
Sembay \etal 2002, Brinkmann \etal 2003, and Ravasio \etal
2004) and earlier extended studies with ASCA and BeppoSAX. In
the following we summarize our main findings which might help
to understand the complex behavior of the source.
   
\subsection{The CCF results}

In general, the observed soft and hard band variations are well
correlated in all three observations.  However, using a
sliding window cross-correlation analysis, we could show that
the cross links between the soft and the hard bands are really
complicated and not easily understood.  There exists a
`characteristic' time scale (of the order of a few ksec) on
which the CCF appears to change ``continuously'' and, depending
on the length of the observing window and the actual activity
state of the source, we find periods with positive or negative
or no lags, but also periods of weak correlations between the
soft and hard energy bands:

\begin{itemize}
\item Periods of weak correlations appear when variations in
the hard band flux or hard band spectra slope are absent in the
soft band.
\item Strong correlation with no obvious lags is observed
throughout orbit 807, although the flux does go up and down. In
fact, this orbit shows the highest amplitude flux variations!
\item In cases where we observe the soft leading the hard band
variations (e.g. P1 and P2 in orbit 546) or the hard leading
the soft (e.g. at $\sim$ 23 ksec in orbit 546) the delays are
just a few minutes in agreement with the lags already
noticed by Brinkmann \etal (2003).
\item The analysis of orbit 84 shows a much more complicated
picture than anticipated from Brinkmann \etal (2003) where the
observation was split in only two parts.  Here we find the hard
flux leading the soft at the beginning, then loss of
correlation, then the soft flux leading, then the hard leading
briefly, and then a loss of correlation in the prolonged
``flare plateau" phase, where the hard band variations are
absent in the soft band. 
\end{itemize}

There is, certainly, a continuous range of variability
time scales, as seen for example in orbit 807, and up to a day
or longer (Kataoka \etal 2001). However, the abovementioned
`characteristic' time scale of correlated emission, i.e., the
time scale over which the correlation between hard and soft
band emission is found to exist, to change sign, or to
disappear, might be representative for the size of the
individual emission regions. With a bulk Lorentz factor of the
jet of $\Gamma \sim 10$ this would correspond to an emission
region of a few$\times10^{15}$cm.

\subsection{The relation between spectral and flux variations}

The cross-correlation analysis demonstrates, on the other 
hand,
the limitations of this method as only little information about
the actual spectral variations can be extracted. We therefore
used the high signal to noise ratio to perform proper spectral
fits to the data down to time scales of $\simgt$ 20 secs.  
Mrk\,421 shows significant spectral variations on time scales
as short as $\sim 500-1000$ sec and most of the time, the hard
and soft band slopes are {\it not} the same. Generally, the
slope of the photon spectrum at hard energies is steeper than
the soft band slope but there are cases when the hard band
slope is {\it flatter} than the soft band slope

The spectral behavior follows the well-known `the
brighter the harder' trend.  Both, the hard and the soft band
spectral indices show a non-linear correlation with the source
flux, and there is an indication that the source may populate
different spectral states at different times in the ``$\Gamma$ 
vs flux" plane. In general, as the flux rises, both the soft
and hard band spectra flatten, but $\Gamma_{soft}$ flattens
faster.  Then the spectra become steeper as the flux decreases,
but now $\Gamma_{hard}$ steepens faster than $\Gamma_{soft}$.
On long time scales $\Gamma_{soft}$ and $\Gamma_{hard}$ vary in
phase and they vary in phase with flux.

On shorter time scales, during the evolution of some flare
like events, we observe ``hysteresis'' phenomena.  
$\Gamma_{soft}, \Gamma_{hard}$ and their ratios follow
loop-like paths in the spectral vs flux plots in the clockwise
(P1 in orbit 546) and anti-clockwise (P1 in orbit 84)
direction. In other cases though, both evolve on a straight
line during rise and decay of the flux (P2 in orbit 84), we
observe a flare where only one follows a loop (P2 in 807), and
there is another case where only the slope ratios follow a well
defined structure (P2 in orbit 546). 

\subsection{Implications to physical models}
   
The numerous instances where the source does not follow the
``average" standard pattern show that any determination of the
physical parameters of the emission region from simple
`homogeneous', `one-zone', etc., models might be questionable.  
There are cases, for example P1 and P2 in orb 546, which could
be "flares" that correspond to emission from a single,
perturbed region of the jet. They show hard lags, so the whole
events are governed by the ``acceleration" time scale of the
relativistic electrons.  Even the soft lags that we see during
the flux drop before P1, could indicate evolution governed by
the cooling time scale. Interestingly, these ``clear" events
show up when the source is at its lowest flux state. It could
be that in these cases we have mainly only one region emitting
at any time. In general, the complex spectral behavior
indicates that the emission geometry is not that simple.  At a
time t$_{\mathrm obs}$ we are not only seeing the emission from
a certain position R(t) in the relativistically moving jet, but
as well the earlier emission from all positions $\sim$ R - c
$\delta t$ down the jet.  In the framework of the commonly
accepted shock-in-jet model this emission is determined by the
history of the colliding shells (blobs) - and, perhaps, by the
earlier (or later) emission of other, independent shells.
 
Even in the framework of the basic, two$-$colliding$-$shells
model (Spada \etal 2001), non-linearities are expected.  Most
of the emission is, certainly, generated when the second,
faster shell hits the first, slower shell. Particles are
accelerated, magnetic fields are amplified and part of the
available energy of this shocked region is radiated. However,
the first shell travels as well with supersonic speeds along
the jet, a shock structure has developed at its front and a
reverse shock travels back through the shell and, after some
time, runs into the shocked emission region of the two
colliding shells. Depending on the initial (and boundary)
conditions of the system this might lead to noticeable changes
of the physical state of the emission region and to
'unexpected' changes of the emission characteristics.
 
\begin{figure}
\psfig{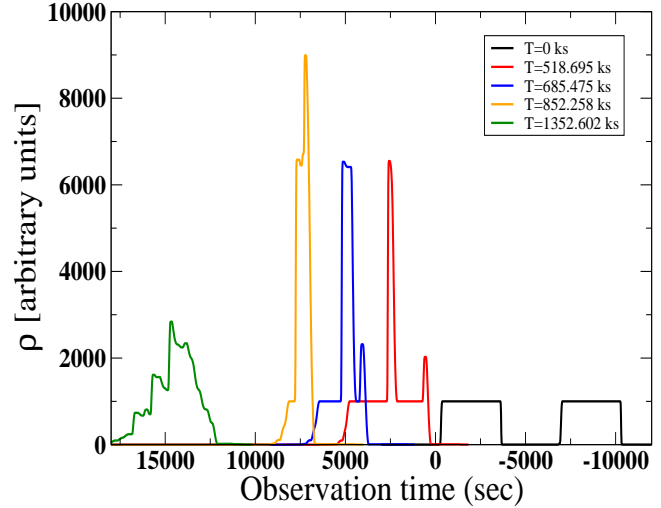}
\caption[]{Density (in arbitrary units) of the colliding shells at different
 evolution times (in the jet frame) of the system as seen by a distant
observer. The observation time (with arbitrary origin) runs from right to
left. The plot shows at what times the observer would `see' the various structures
 approaching him with relativistic velocities. }
\label{figure:petar1}
\end{figure}
 
\begin{figure}
\psfig{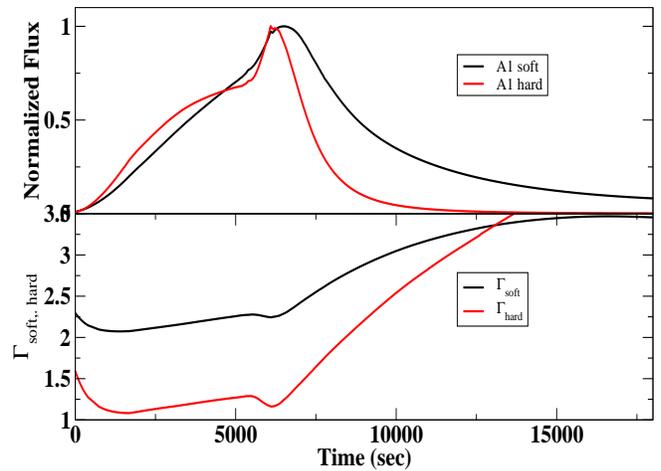}
\caption[]{ Normalized X-ray emission from the model of Fig. \ref{figure:petar1}
in the top panel in the hard (1$-$10 keV, red curve) and the soft (0.1$-$1 keV) X-ray band.
The bottom panel shows the evolution of the spectral power law indices in the
two bands.}   
\label{figure:petar2}
\end{figure}

These effects are illustrated in  the following figures, obtained from numerical 
relativistic hydrodynamic simulations of two colliding shells with
Lorentz factors $\Gamma_1 = 5$ and $\Gamma_2 = 7.5$ (Mimica \etal 2004a, 2004b).
Fig. \ref{figure:petar1} shows the density distribution of the shells at five
different times in the jet frame as seen by a distant observer.
 The origin of the observer's
time scale is arbitrarily chosen. The configuration at T = 0 in the jet frame
(black curves)  would be seen by a distant observer between  
-10000 $\leq t_{\mathrm obs} \leq 0 $ s as two individual shells moving towards 
him.
The situation at T $\sim$ 519 ksec in the jet frame (red curve) would first appear
to the observer as the shock structure at the front of shell 1 and then, around
t$_{\mathrm obs} \sim $ 2500 sec, as strongly rising emission as the two shells have
collided.  From then on, up to about  t$_{\mathrm obs} \sim$ 5000 sec, the  
emission would be rising due to the growing density and size of the collision
region in the jet frame.  At t$_{\mathrm obs} \sim$ 7500 sec, at the 
evolutionary time  T $\sim$ 852 ksec in the jet frame  (yellow curve),
 the shock structure from
shell 1 has merged with the interaction region  of the shells,
changing strongly the emission properties. Then, at later times, the structure
fades away.  
  
Fig. \ref{figure:petar2} shows the normalized light curve of this event 
in the 0.1$-$1 keV and the 1$-$10 keV X-ray band in the upper panel and the
evolution of the spectral indices in the lower panel. 
It must be emphasized that
the parameters chosen for this simulation are not at all applicable for the
physical conditions in Mrk\,421. The spectra  indicate that the shocks are too
strong  and/or the magnetic field amplification has been over-estimated;
they are just presented for illustrative purposes.
In particular, the spectral hardening phase occurs only at the very beginning
of the rising light curve, then the spectrum starts cooling. At the time
where the reverse shock from the first shell joins the inter-shell emission 
region, we find spectral and intensity changes of the emission which would 
not fit into a 'regular' evolution pattern of the observed flare and the 
cross correlation analysis of this light curve would predict lags of 
changing sign and amplitude. Looking at the above presented observations it appears
that, for example, parts of the light curve of orbit 807 (Fig. \ref{figure:807gam})
 resemble this 
behavior: the flux and spectral index of the soft energy band varies smoothly while
the flux and power law index in the hard band exhibits some `unexpected' deviations. 
 
One of the most interesting questions raised by the observed emission is  
the cause of the long-term intensity changes. In the three observations
analyzed here, the flux changed by a factor of more than three. 
Are these variations caused
by  geometrical changes of the viewing conditions, like in the  lighthouse
model by Camenzind \& Krockenberger (1992)? 
Or are we seeing periods of enhanced activity, a dramatic increase of the
number of interacting blobs or stronger interactions - for which the
individual emission patterns are averaged and form an
enhanced 'background' emission.   
There is growing evidence that the blazar emission 
consists of a `quasi-stationary' component which changes only on
long time scales and a flaring component which gives rise to the short
time scale intensity and spectral variations. The influence of these different
components on the determination of the physical parameters deduced 
from the observations has been discussed in more detail by Fossati \etal (2000a)
 and Fossati \etal (2000b).
In extensive numerical simulations of the shock-in-jet 
scenario, varying the injection rate, the duration, speeds and other 
relevant physical parameters of the shells, we could not reproduce the 
observed variability pattern of the emission of Mrk\,421. Only after 
adding a substantial 'background' emission of $\simgt$ 60\%  to the
numerically estimated flux we did find light curves that looked similar to the
observed ones.

\section{Conclusions}
We have presented a  time resolved analysis of the currently available XMM-Newton
observations of Mrk~421 performed with the PN in timing mode.
The high signal to noise ratio allowed us to 
perform cross-correlation analyses  between the soft and hard energy bands
with a sliding window technique
which revealed that the characteristics of the correlated emission changes
on time scales of a few ksec.
With time-resolved spectral fits on time scales of a few hundred seconds
we could study the spectral evolution of the source. 
We find significant spectral variations on time scales
as short as $\sim 500-1000$ sec and non-linear correlations between between
slopes of the fitted broken power law model and the flux of the source.
 
The temporal and spectral behavior of the source is very complex and during
flares various variability patterns in the soft and hard energy band
were observed.
Correspondingly, it appears hard to deduce uniquely the underlying physical
parameters for the emission process from the observations.
We compared the observations with relativistic hydrodynamic simulations 
of the currently favored 'shock-in-jet' model for the BL Lac emission
(see, for example, Spada et al. 2001) which takes into account that 
we are seeing the 
emission from multiple shocks which have either very different
physical parameters or that we detect the emission from similar
shocks at very different states of their evolution, heavily confused
by relativistic beaming and time dilatation effects.
As the actual emission characteristics of these numerical models depend
 on a large number
of  physical parameters  and poorly known boundary conditions it is
a tedious task to reproduce numerically
the large variety of observed light curves.  However, BL Lac jets are
relatively  simple physical systems, which might be solely
governed by the equations of relativistic MHD and special relativity.
Thus, further observational progress (in particular long continuous
\xmm observations  with high spectral and
temporal resolution) and extended numerical simulations might finally
lead to a better understanding of the physical conditions in these systems.

\vskip 0.4cm
\begin{acknowledgements}
This work is based on observations with XMM-Newton, an ESA science mission 
with instruments and contributions directly funded by ESA Member States
 and the USA (NASA).
\end{acknowledgements}

\end{document}